\documentclass[twocolumn]{aastex701}
\received{2025 July 2}
\revised{2025 July 30}
\accepted{2025 August 14} 
\shorttitle{{\sc 21cmKAN}: Transparent Global 21 cm Signal Emulation}
\shortauthors{Dorigo Jones et al.}
\usepackage{newtxtext, newtxmath, anyfontsize, threeparttable, booktabs, multirow}
\newcommand{\RomanNumeralCaps}[1]{\MakeUppercase{\romannumeral #1}}
\makeatletter
\newcommand\footnoteref[1]{\protected@xdef\@thefnmark{\ref{#1}}\@footnotemark}
\makeatother

\begin{document}
\title{The Wrath of KAN: Enabling Fast, Accurate, and Transparent Emulation of the Global 21 cm Cosmology Signal}
\correspondingauthor{J. Dorigo Jones}
\author[0000-0002-3292-9784]{J. Dorigo Jones}
\affiliation{Center for Astrophysics and Space Astronomy, Department of Astrophysical and Planetary Sciences, University of Colorado Boulder, CO 80309, USA}
\email[show]{johnny.dorigojones@colorado.edu}
\author[0000-0003-3496-7490]{B. Reyes}
\affiliation{University of Colorado Research Computing, University of Colorado Boulder, CO 80309, USA}
\email{brandon.reyes-2@colorado.edu}
\author[0000-0003-2196-6675]{D. Rapetti}
\affiliation{NASA Ames Research Center, Moffett Field, CA 94035, USA} \affiliation{Research Institute for Advanced Computer Science, Universities Space Research Association, Washington, DC 20024, USA}
\affiliation{Center for Astrophysics and Space Astronomy, Department of Astrophysical and Planetary Sciences, University of Colorado Boulder, CO 80309, USA} 
\email{david.rapetti@colorado.edu}
\author[0000-0003-0016-5377]{Shah Mohammad Bahauddin}
\affiliation{Laboratory for Atmospheric and Space Physics, University of Colorado, Boulder, CO 80303, USA}
\affiliation{Center for Astronomy, Space Science and Astrophysics, Independent University, Bangladesh, Dhaka 1229, Bangladesh}
\email{Shah.Bahauddin@lasp.colorado.edu}
\author[0000-0002-4468-2117]{J. O. Burns}
\affiliation{Center for Astrophysics and Space Astronomy, Department of Astrophysical and Planetary Sciences, University of Colorado Boulder, CO 80309, USA} 
\email{jack.burns@colorado.edu}
\author[0009-0002-4218-4840]{D. W. Barker}
\affiliation{Center for Astrophysics and Space Astronomy, Department of Astrophysical and Planetary Sciences, University of Colorado Boulder, CO 80309, USA} 
\email{david.w.barker@colorado.edu}

\begin{abstract} 
Based on the Kolmogorov-Arnold Network (KAN), we present a novel emulator of the global 21 cm cosmology signal, {\sc 21cmKAN}, that provides extremely fast training speed while achieving nearly equivalent accuracy to the most accurate emulator to date, {\sc 21cmLSTM}. The combination of enhanced speed and accuracy facilitated by {\sc 21cmKAN} enables rapid and highly accurate physical parameter estimation analyses of multiple 21 cm models, which is needed to fully characterize the complex feature space across models and produce robust constraints on the early universe. Rather than using static functions to model complex relationships like traditional fully-connected neural networks do, KANs learn expressive transformations that can perform significantly better for low-dimensional physical problems. {\sc 21cmKAN} predicts a given signal for two well-known models in the community in 3.7 ms on average and trains about 75 times faster than {\sc 21cmLSTM}, when utilizing the same typical GPU. {\sc 21cmKAN} is able to achieve these speeds because of its learnable, data-driven transformations and its relatively small number of trainable parameters compared to a memory-based emulator. We show that {\sc 21cmKAN} required less than 30 minutes to train and fit these simulated signals and obtain unbiased posterior distributions. We find that the transparent architecture of {\sc 21cmKAN} allows us to conveniently interpret and further validate its emulation results in terms of the sensitivity of the 21 cm signal to each physical parameter. This work demonstrates the effectiveness of KANs and their ability to more quickly and accurately mimic expensive physical simulations in comparison to other types of neural networks.
\end{abstract}

\section{Introduction} \label{sec:intro}
The physics of the early Universe is imprinted on the redshifted 21 cm cosmological signal emitted by neutral hydrogen (HI). Thus, measuring its brightness temperature relative to the radiation background may reveal key properties of the first stars and galaxies and the intergalactic medium during the Cosmic Dawn and Dark Ages (see \citealt{Furlanetto06} for a review). Robust software and modeling tools are needed to achieve these constraints, in particular efficient and accurate simulations of the 21 cm signal that can be used to perform Bayesian inference and estimate the underlying physical parameters that shape its evolution. Seminumerical or semianalytical cosmological simulations of the 21 cm signal (e.g., \citealt{Mesinger11, Fialkov14, Mirocha17}) offer faster alternatives to hydrodynamic codes that solve the governing partial differential equations (PDEs); however, it is computationally prohibitive to use these models directly in inference pipelines. This has motivated the development of emulators of the 21 cm signal -- based on deep artificial neural networks (NNs) -- that mimic the output of cosmological simulations for given sets of input physical parameters in just milliseconds. Replacing computationally expensive models with emulators is essential to efficiently sample parameter posteriors from a measured signal (e.g., \citealt{REACH, Bevins24}), assuming systematic effects are properly addressed (e.g., \citealt{PaperII, Anstey23, Saxena23a}).

A summary statistic of particular interest is the 1D isotropic (i.e., global; \citealt{Shaver99}) 21 cm signal, which is observed at low radio frequencies of $\nu \lesssim 235$ MHz, corresponding to redshifts $z\gtrsim5$. Existing emulators of the global 21 cm signal (\citealt{Cohen20, globalemu, 21cmVAE, Breitman24, DorigoJones24}) differ in how they prioritize training speed, evaluation speed, and emulation accuracy. In this paper, we present {\sc 21cmKAN} -- a new publicly available global 21 cm signal emulator utilizing a Kolmogorov-Arnold Network (KAN\footnote[7]{\label{KAN}\url{https://github.com/KindXiaoming/pykan}}; \citealt{KAN}) that trains extremely fast and achieves low emulation error comparable to the most accurate known emulator, {\sc 21cmLSTM} \citep{DorigoJones24}. It is beneficial to have an emulator that is very fast to train, while maintaining high accuracy similar to {\sc 21cmLSTM}, to emulate different models of the global 21 cm signal and constrain their cosmological and astrophysical parameters. For upcoming observations, the speed--accuracy combination of {\sc 21cmKAN} eliminates emulator training as a bottleneck by enabling rapid and accurate posterior sampling to comprehensively explore signal models.

KANs are a recently introduced NN architecture inspired by the Kolmogorov-Arnold representation theorem (\citealt{kolmogorov}; see Section~\ref{subsec:KAN}). KANs learn complex relationships between input parameters and the desired output in a highly adaptive and expressive manner, by stacking layers of data-driven functional transformations (see Figure~\ref{fig:KAN}). In contrast, traditional fully-connected NNs \citep{Rumelhart86} transform input to output using fixed, pre-determined functions with learned scalar weights. The flexibility of KANs makes them a promising alternative to traditional NNs and particularly well-suited to modeling complex, lower-dimensional physical systems like the global 21 cm signal (see also \citealt{Cui25, Shi25}). KANs provide an inherently transparent architecture with relatively few parameters that allows the user to see what it learns, which is desired in scientific applications of machine learning to further validate and explain predictions and enable transferability of knowledge between models \citep{Montavon18, Siemiginowska19, Wang23, Belfiore25}.

To demonstrate the benefits of using KANs to emulate the global 21 cm signal, the paper is organized as follows. In Section~\ref{sec:methods}, we describe the architecture and training of {\sc 21cmKAN}. In Section~\ref{sec:results}, we present the emulation accuracy and speed of {\sc 21cmKAN} compared to other emulators and posterior constraints when using {\sc 21cmKAN} to fit mock 21 cm signals, and we show examples of and interpret the learned activations. Finally, we summarize the conclusions in Section~\ref{sec:conclusions}.

\section{Methods} \label{sec:methods}
\subsection{Neural Networks and the Architectural Benefits of KANs} \label{subsec:KAN}
\begin{figure*}
    \includegraphics[scale=0.65]{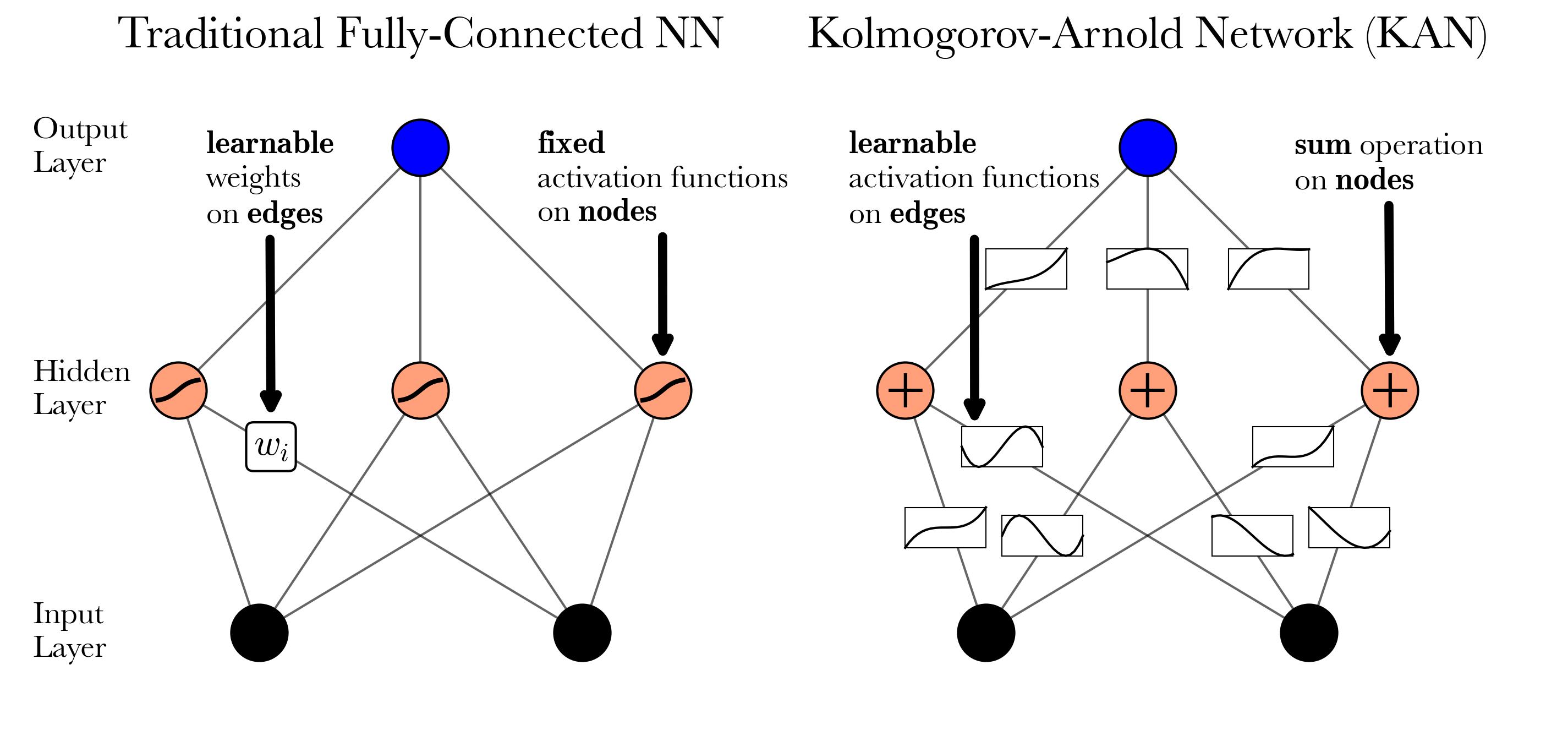}
    \centering
    \caption{Architecture diagrams of a single-hidden-layer traditional fully-connected NN (left) and Kolmogorov-Arnold Network (right). In the KAN, activation functions are learned and applied to parameters on the edge connections between nodes and summed at the nodes. For traditional fully-connected NNs, the activations are pre-determined and fixed on the nodes, the scalar weights of which are learned on the edges.\label{fig:KAN}}
\end{figure*}

\begin{figure*}
    \includegraphics[scale=0.32]{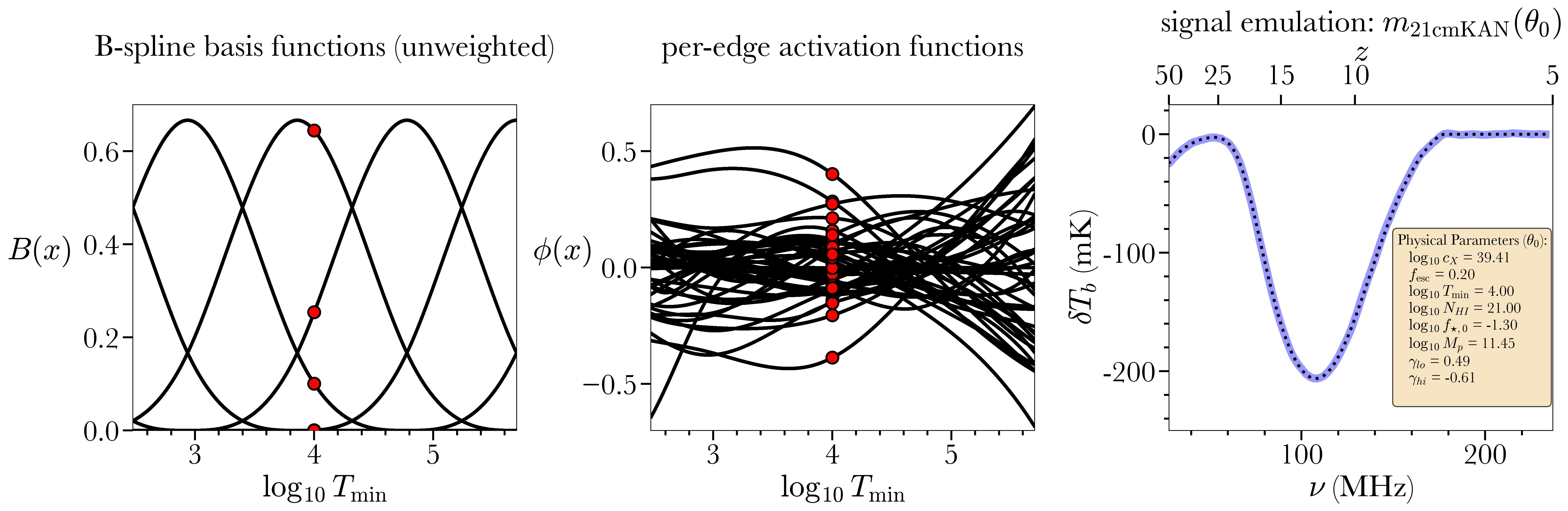}
    \caption{Example learned components and signal emulation of \textsc{21cmKAN} when trained on the physical model \texttt{ARES}. \textit{Left:} Unweighted B-spline basis functions, $B(x)$, that parameterize the trainable activation functions, $\phi(x)$. The basis functions are evaluated at a sample input (e.g., $\log_{10} T_{\rm min} = 4$, red markers) and linearly combined to yield the activation functions. \textit{Middle:} Resulting per-edge activation functions, $\phi(x)$, each learned independently along the network edges (see Figure~\ref{fig:KAN}). \textit{Right:} Emulated global 21 cm brightness temperature, $\delta T_b(\nu)$, as a function of frequency $\nu$, in blue, output by \textsc{21cmKAN} for input astrophysical parameters ${\theta}_0$ (see Table~\ref{tab:params} for parameter descriptions). The corresponding ``true'' signal from {\tt ARES} is shown in dotted black. See Sections~\ref{subsec:data},~\ref{subsec:architecture}, and~\ref{subsec:model_comparison} for further details on the data, training, architecture, and interpretation of {\sc 21cmKAN}.\label{fig:full}}
\end{figure*}

At the core of a NN lies the activation function, inspired by the firing behavior of biological neurons. Each individual neuron -- or ``node'' in a NN -- processes the weighted sum of incoming signals by applying a nonlinear function, which determines whether and to what extent the node activates. This nonlinear response allows each node to contribute selectively to the network’s representation of complex input patterns, enabling the construction of rich hierarchical features across layers. By performing a series of these simple mathematical transformations, NNs become powerful function approximators, as established by the universal approximation theorem \citep{Cybenko, Hornik}.

A traditional fully-connected NN, shown schematically in the left panel of Figure~\ref{fig:KAN}, is composed of layers of interconnected nodes, where each edge connection between nodes carries a scalar weight, $w_i$, and each node applies a fixed activation function (e.g., $\tanh$) to a weighted sum of its inputs plus a constant term (known as bias). In this framework, the trainable parameters consist of the weights and biases, while the activation functions remain fixed throughout training. The expressive power of a traditional fully-connected NN -- also referred to as a multilayer perceptron (MLP) -- arises from the composition of many simple nonlinear transformations. However, because these transformations are distributed across many layers and involve fixed nonlinearities, the resulting model is often difficult to interpret and lacks transparency in how it represents the underlying functional relationships \citep{Montavon18, Lipton18, Rudin18}, although there are methods to help explain the predictions and behavior of so-called ``black box'' machine learning models (e.g., \citealt{Simonyan14, Koh17, Kim18}).

In contrast to traditional NNs, KANs assign trainable functions to the edges, as illustrated in the right panel of Figure~\ref{fig:KAN}, while the nodes sum the transformed inputs. Each function, $\phi(x)$, is learned through a parameterized B-spline: a class of smooth, piecewise polynomial curves commonly used in approximation theory. Basis functions, $B(x)$, are weighted to form the B-splines at each edge that ultimately determine the final network output, a process that we depict in Figure~\ref{fig:full} (see Section~\ref{subsec:model_comparison} for further detail). The KAN architecture is motivated by the Kolmogorov–Arnold representation theorem (\citealt{kolmogorov}), which states that any multivariate continuous function can be written as the finite composition of univariate continuous functions. By learning these univariate transformations and their summations, KANs can capture functional relationships more efficiently. As a result, training is often faster and convergence is robust, particularly in low-dimensional problems involving functional compositions, which are frequently encountered in astrophysics and cosmology. Since each transformation is explicitly represented as a learned function, the internal representations in KANs are easier to interpret and analyze when compared to traditional fully-connected NNs \citep{KAN}.

In prior work, we showed that the {\sc 21cmLSTM} model provided the most accurate emulations of the global 21 cm signal to date \citep{DorigoJones24}, owing to its inherent capability to capture the temporal evolution of the 21 cm brightness temperature across redshifts or low radio frequencies. Recurrent architectures, such as Long Short-Term Memory (LSTM) networks, are specifically designed to model sequential data. They achieve this by learning how to store and propagate information through time (or sequence) using gated mechanisms (e.g., input, forget, and output gates) that allow them to retain memories and capture temporal dependencies in data. However, LSTM NNs can be slower to train and often require more complex optimization.

In contrast to LSTM NNs, KANs capture the shape of the signal by directly learning the underlying functional transformations, without the need for memory gates or recurrent structures. This makes them well suited for low-dimensional, multivariate curve emulation where preserving the sequential structure is important. We have thus created the {\sc 21cmKAN} emulator to leverage the functionality of KANs for the global 21 cm signal. As demonstrated in Section~\ref{sec:results}, while {\sc 21cmLSTM} achieved the highest overall accuracy among conventional neural architectures, {\sc 21cmKAN} attains comparable performance with significantly faster convergence and reduced architectural complexity.

The transparent nature of the KAN architecture, illustrated by its decomposable components in Figures~\ref{fig:KAN} and~\ref{fig:full}, allows the user to directly visualize the functional transformations learned by {\sc 21cmKAN} and conveniently infer feature importance (see Section~\ref{subsec:model_comparison}). This builds further trust and reliability in the model's predictions beyond the reported emulation error and complements explainable machine learning techniques, which are benefits of interpretable architectures \citep{Lipton18, KAN}. Additionally, KANs are more geared towards transferability of knowledge between simulations or parameterizations (i.e., domain adaptation or transfer learning) because of their simplicity and flexibility compared to traditional NNs. Such functionality would help make global 21 cm signal inference pipelines more robust, generalizable, and model agnostic by unifying physical parameters across simulations and reducing the necessary sizes of training sets (e.g., \citealt{Wehenkel25, Andrianomena25, Belfiore25}).

{\sc 21cmKAN} utilizes Ef-KAN\footnote[8]{\label{EfKAN}\url{https://github.com/Blealtan/efficient-kan}}, which is an efficient implementation of KANs that significantly reduces the memory and evaluation cost, in comparison to the original implementation\footnoteref{KAN} \citep{KAN}, by reformulating the activation execution. {\sc 21cmKAN}\footnote[9]{\label{code}\url{https://github.com/jdorigojones/21cmKAN}} is written in {\sc Python} with a {\sc PyTorch} \citep{pytorch} backend and is publicly available. These attributes make {\sc 21cmKAN} simple to train, test, and apply to different models and data sets. In the following subsections, we describe how the {\sc 21cmKAN} emulator can be trained, using popular and publicly available data sets.

\subsection{Data} \label{subsec:data}
To train and test {\sc 21cmKAN}, we utilize publicly available data sets\footnote[10]{\label{data}\dataset[doi: 10.5281/zenodo.5084113]{https://doi.org/10.5281/zenodo.5084113}; \dataset[doi: 10.5281/zenodo.13840725]{https://doi.org/10.5281/zenodo.13840725}} of synthetic global 21 cm signals that allow for direct comparison of speed and accuracy against existing emulators that used the same data sets. The ``{\sc 21cmGEM}'' set was made from a large-volume seminumerical model \citep{Visbal12, Fialkov13, Fialkov14} similar to {\sc 21cmFAST} (\citealt{Mesinger11}; see \citealt{Cohen20}), while the ``{\tt ARES}'' set was created by a physically-motivated, semianalytical model called Accelerated Reionization Era Simulations ({\tt ARES}\footnote[11]{\url{https://github.com/mirochaj/ares}}; \citealt{Mirocha12, Mirocha14, Mirocha17}) that does not compute 3D volumes. See Figure 2 of \citet{DorigoJones24} for representative subsets of each model data set. By training and testing on these data sets, we are able to make direct comparisons of {\sc 21cmKAN} with the emulators {\sc 21cmGEM} \citep{Cohen20}, {\tt globalemu} \citep{globalemu}, {\tt 21cmVAE} \citep{21cmVAE}, and {\sc 21cmLSTM} \citep{DorigoJones24}.

Although the two sets were made using different physical models with different parameters, the {\tt ARES} set was made to be nearly physically equivalent to the {\sc 21cmGEM} set (see \citealt{DorigoJones24}). Seven astrophysical parameters were varied over wide ranges to create the {\sc 21cmGEM} set, while eight parameters were varied to create the {\tt ARES} set (see Table~\ref{tab:params}). The parameters in each set control the star formation efficiency (SFE) and ionizing photon production in galaxies, although via different parameterizations. The signals in the {\sc 21cmGEM} set span $z=5-50$, while the {\tt ARES} signals span $z=5.1-49.9$, all with resolution of $\Delta z=0.1$. Each test set is the same size (1704), and the combined training+validation sets are similar in size to within 3\% (27,292 for {\sc 21cmGEM} and 26,552 for {\tt ARES}, each split 90\% for training and 10\% for validation). Each set has a physical constraint on the HI fraction ($x_{\rm HI}$) at the end of the Epoch of Reionization, motivated by observations (see, e.g., \citealt{Fan06, McGreer15, Bosman22, Jin23}): the {\sc 21cmGEM} set requires $x_{\rm HI}<$16\% at $z=5.9$, and the {\tt ARES} set requires $x_{\rm HI}<$5\% at $z=5.3$.

Preprocessing or normalization of the data (i.e., physical parameter values) and labels (i.e., signal brightness temperatures, $\delta T_b$) is a common step before network training to facilitate performance. To preprocess the data, we first take the $\log_{10}$ of those parameters that are uniform only in $\log_{10}$-space: $f_{\star}$, $V_c$, and $f_X$ for {\sc 21cmGEM} and $c_X$, $T_{\rm min}$, $f_{\rm \star,0}$, and $M_{\rm p}$ for {\tt ARES} (see definitions of parameters in Table~\ref{tab:params}). The labels are flipped to train the network from high-$z$ to low-$z$. Finally, a global (rather than bin-by-bin) Min-Max normalization is performed on each feature, $x$, in the data and labels:
\begin{equation} \label{eqn:minmax}
\tilde{x} = \frac{x-x_{\rm min}}{x_{\rm max}-x_{\rm min}}.
\end{equation}

\begin{table*}
    \caption{Astrophysical Parameters Varied in {\sc 21cmGEM} and {\tt ARES} Data Sets and Fit in Nested Sampling Analyses\label{tab:params}}
    \begin{center}
    \begin{tabular}{cccc}
    \toprule
    Model & Parameter & Description & Range (with units)\\
    \hline
    \multirow{7}{*}{{\sc 21cmGEM}} & $f_*$ & star formation efficiency & Log unif. [$10^{-4}$, $5\times10^{-1}$]\\
    & $V_c$ & minimum circular velocity of star-forming halos & Log unif. [$4.2$, $100$] km s$^{-1}$\\
    & $f_X$ & X-ray efficiency of sources & Log unif. [$10^{-6}$, $10^3$]\\
    & $\tau$ & cosmic microwave background (CMB) optical depth & Uniform [$0.04$, $0.2$]\\
    & $\alpha$ & slope of X-ray spectral energy distribution (SED) & Discrete [$1$, $1.5$]\\
    & $\nu_{\rm min}$& low energy cut off of X-ray SED & Discrete [$0.1$, $3$] keV\\
    & $R_{\rm mfp}$& mean free path of ionizing radiation &  Uniform [$10$, $50$] Mpc\\    
    \cline{1-4}
    \multirow{8}{*}{{\tt ARES}} & $f_{\rm \star,0}$ & peak star formation efficiency & Log unif. [$10^{-5}$, $10^0$] \\
    & $T_{\rm min}$ & minimum temperature of star-forming halos & Log unif. [$3\times 10^2$, $5\times 10^5$] K \\
    & $c_X$ & normalization of X-ray luminosity--star formation rate relation & Log unif. [$10^{36}$, $10^{44}$] erg s$^{-1}$($M_{\odot}$ yr$^{-1}$)$^{-1}$\\
    & $\log N_{\rm H \RomanNumeralCaps{1}}$ & neutral hydrogen column density in galaxies & Uniform [18, 23] \\
    & $M_{\rm p}$ & dark matter halo mass at $f_{\rm \star,0}$ & Log unif. [$10^8$, $10^{15}$] $M_{\odot}$ \\
    & $\gamma_{\rm lo}$ & low-mass slope of $f_{\rm \star} (M_{\rm h})$& Uniform [0, 2] \\
    & $\gamma_{\rm hi}$ & high-mass slope of $f_{\rm \star} (M_{\rm h})$ & Uniform [-4, 0] \\
    & $f_{\rm esc}$ & escape fraction of ionizing radiation & Uniform [0, 1] \\
    \bottomrule
    \end{tabular}
    \end{center}
\end{table*}

\subsection{Training and Architecture} \label{subsec:architecture}
\begin{figure}
    \includegraphics[scale=0.47]{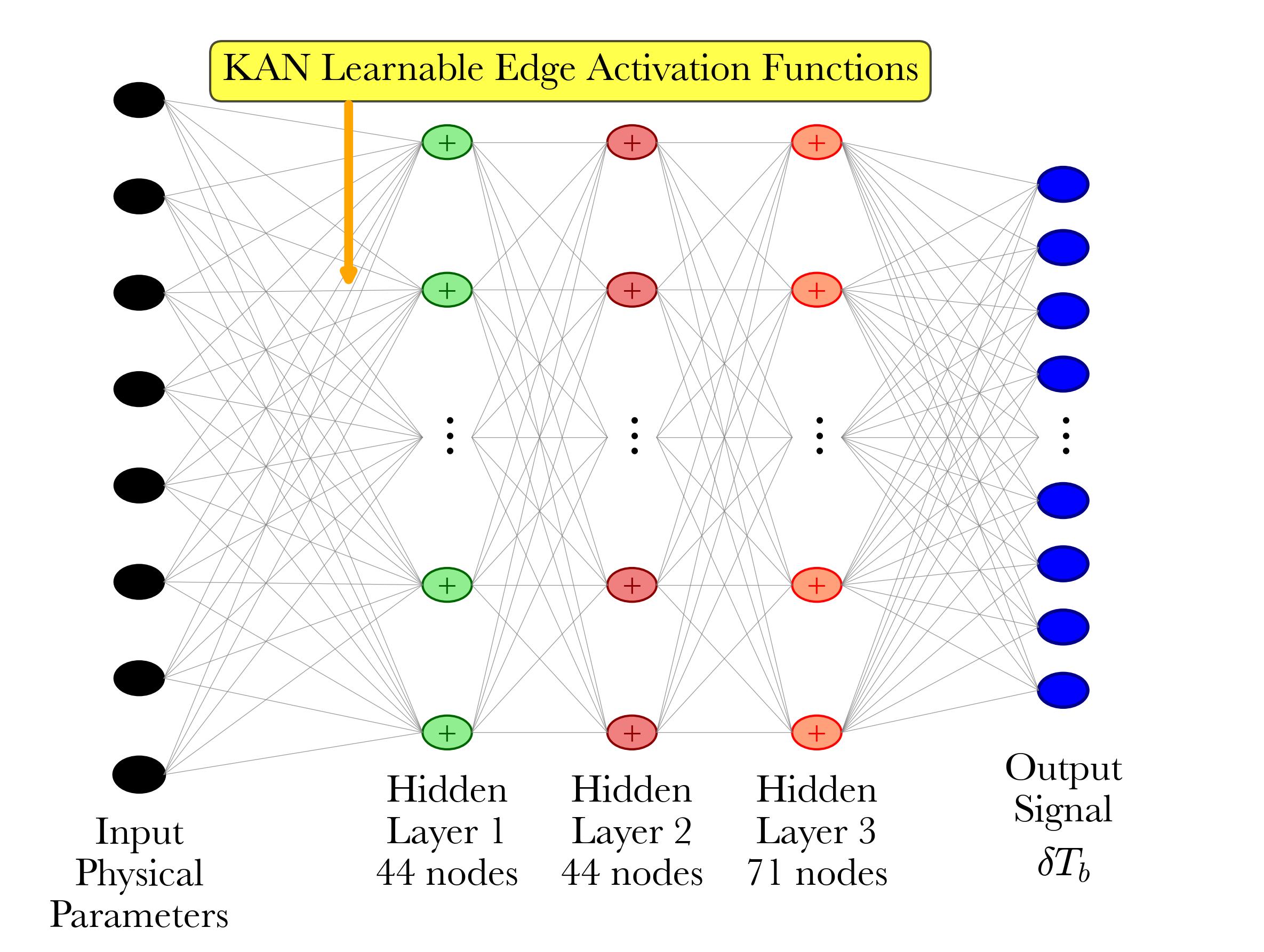}
    \centering
    \caption{Architecture diagram of {\sc 21cmKAN}. As described in Section~\ref{subsec:KAN}, {\sc 21cmKAN} learns and applies flexible activation functions on the edges and sums them at the nodes to accurately and efficiently map the input physical parameters to the output global 21 cm signal. See Section~\ref{subsec:architecture} for a detailed description of the architecture, including optimization of the numbers of hidden layer nodes, parameterizations of the B-spline activations, and network training.\label{fig:schematic}}
\end{figure}

For results presented in this work, we trained and tested {\sc 21cmKAN} using a single NVIDIA A100 GPU with 10 CPU cores on the Alpine heterogeneous supercomputing cluster operated by University of Colorado Boulder Research Computing (CU RC; \citealt{Alpine}). The emulator is trained on the preprocessed training set and for each complete pass of the training set through the network (i.e., for each epoch) saves the mean squared error (MSE; Equation~\ref{eqn:mse}) loss function values between the true and emulated normalized signals in the training and validation sets:
\begin{equation} \label{eqn:mse}
{\rm MSE} = \langle(\delta T_{\rm b, true}(\nu)-\delta T_{\rm b, emulated}(\nu))^2\rangle.
\end{equation}
\noindent The training set errors are used during backpropagation to calculate the loss gradients and update the weights via stochastic gradient descent, which is optimized using the Adam method \citep{Adam} with default learning rate of 10$^{-3}$ and weight decay of 10$^{-4}$. The validation set is used to monitor for overfitting, ensuring the emulator can generalize to unseen signals. Finally, the test set is used to compute the reported emulation error of the saved trained instance of {\sc 21cmKAN}, which loads the weights from the final epoch. As described below, we optimized the training and architecture of {\sc 21cmKAN} to achieve the smallest reproducible emulation error and a validation loss that plateaus in the final epochs; as a result, we did not find it necessary to implement additional network convergence strategies such as learning rate decay, batch size scheduling, or early stopping.

We performed two separate Tune experiments (\citealt{Tune}) using Optuna \citep{optuna}, which is an automatic hyperparameter optimization software framework designed for machine learning. For these hyperparameter searches, we utilized the NVIDIA Grace Hopper (GH200) superchip through CU RC, which allowed us to train multiple models in parallel. Automatic hyperparameter searches are a common practice when designing NNs (e.g., \citealt{21cmVAE}) and in our case do not significantly affect the resulting emulation accuracy of {\sc 21cmKAN}. The hyperparameters searched were the components of the KAN architecture and the training batch size. The batch size corresponds to the number of signals whose loss values are averaged at a time to update the network weights during each epoch. Specifically, the architecture hyperparameters optimized were the number of nodes in each hidden layer, the number of grid intervals in the B-spline activation functions (i.e., the number of knots, or grid points, minus one; see Section~\ref{subsec:model_comparison}), and the order of the individual splines. Each search trained 1,000 networks on the {\sc 21cmGEM} set for 400 epochs and varied each network's hyperparameter values over the following wide ranges to find nearly optimal combinations: (1) number of nodes in hidden layer one: [10, 100], (2) number of nodes in hidden layer two: [10, 100], (3) number of nodes in hidden layer three: [10, 100], (4) grid intervals: [3, 15], (5) spline order: [3,7], and (6) batch size: [10, 1000]. We manually tested using one, two, three, or four hidden layers, rather than including the number of layers as a searched hyperparameter, and we found that three hidden layers resulted in the most accurate trained network on average.

Based on these hyperparameter searches, we determined a default architecture that achieves the lowest emulation error and has low model complexity (i.e., low number of B-spline basis functions; see Section~\ref{subsec:model_comparison}) to facilitate fast training. We trained four different nearly-optimal network architectures on the {\sc 21cmGEM} set and repeated each for 20 trials to assess the stochasticity in the training algorithm (see Appendix~\ref{sec:diffarchs}; Figure~\ref{fig:appendix_diffarchs}). We evaluated each trained network at the parameter values of the signals in the test set and compared the resulting emulated signals to their corresponding ``true'' signals, computing for each signal the relative rms error in percent across the full frequency range:
\begin{equation} \label{eqn:rel_error}
{\rm Error} = \frac{\sqrt{{\rm MSE}}}{{\rm max}(|\delta T_b(\nu)|)},
\end{equation}
\noindent where MSE is defined by Equation~\ref{eqn:mse} and max($|\delta T_b(\nu)|$) is the signal amplitude.

We found that the three-hidden-layer KAN generally prefers $<75$ nodes per hidden layer, which is much less than what has been found for traditional fully-connected NN emulators of the 21 cm signal (see \citealt{21cmVAE}). This finding is consistent with the results of \citet{KAN} and further highlights the transparency of the KAN architecture. The default architecture of {\sc 21cmKAN} is three hidden layers of 44, 44, and 71 nodes, respectively, and the B-splines are parameterized by seven grid intervals and individual splines of order three (i.e., cubic splines, which is recommended by \citealt{KAN}). The default batch size is 100 signals for each training epoch. We use the same default architecture when training {\sc 21cmKAN} on either {\sc 21cmGEM} or {\tt ARES}; however, we train on the {\sc 21cmGEM} set for 400 epochs and on the {\tt ARES} set for 800 epochs to allow the validation loss to reach a stable value at the final epochs that is indicative of sufficient training without overfitting (see Appendix~\ref{sec:loss}; Figure~\ref{fig:appendix_loss}; Section~\ref{subsec:model_comparison}). The number of input layer nodes is the number of physical parameters for the data set being trained or tested on: seven for the {\sc 21cmGEM} set and eight for the {\tt ARES} set. The number of output layer nodes is the number of frequencies that compose each signal: 451 for the {\sc 21cmGEM} set and 449 for the {\tt ARES} set. Figure~\ref{fig:schematic} depicts the default architecture of {\sc 21cmKAN}.

\section{Results} \label{sec:results}
\subsection{Accuracy} \label{subsec:accuracy}
\begin{figure}
    \includegraphics[scale=0.59]{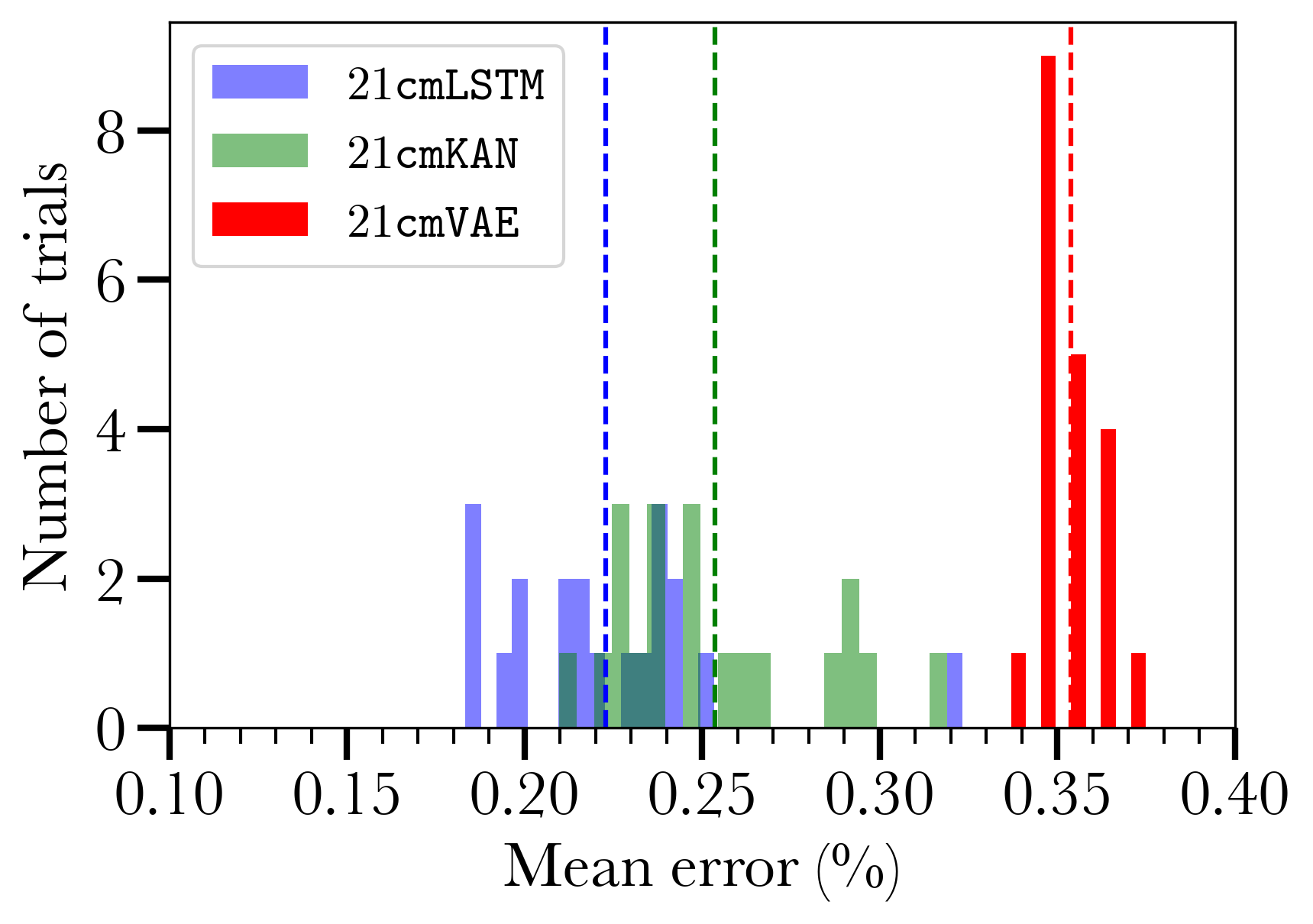}
    \caption{Histogram of the mean relative rms error (Equation~\ref{eqn:rel_error}) for 20 trials of {\sc 21cmKAN} (in green) trained and tested on the {\sc 21cmGEM} data set. The blue histogram is the error for 20 trials of {\sc 21cmLSTM} trained and tested on the same data \citep{DorigoJones24}. The red histogram is the approximate error for 20 trials of {\sc 21cmVAE} trained and tested on the same data (adapted from Figure 6 of \citealt{21cmVAE}). Dashed lines depict the average emulation errors.\label{fig:err_hist}}
\end{figure}

We report the emulation accuracy of {\sc 21cmKAN} when trained and tested on the {\sc 21cmGEM} set, as well as on the {\tt ARES} set (see Section~\ref{subsec:data}), to directly compare to the previous emulators {\sc 21cmGEM} \citep{Cohen20}, {\tt globalemu} \citep{globalemu}, {\sc 21cmVAE} \citep{21cmVAE}, and {\sc 21cmLSTM} \citep{DorigoJones24}. We used the network architecture and training settings described in Section~\ref{subsec:architecture} and trained and tested 20 identical trials of {\sc 21cmKAN} on each set, computing the relative (Equation~\ref{eqn:rel_error}) and absolute (in mK) rms emulation errors.

When trained and tested on the {\sc 21cmGEM} set, the distribution of mean relative rms error for all 20 trials is shown in Figure~\ref{fig:err_hist}. Across the 20 trials, {\sc 21cmKAN} has an average relative mean error of 0.254 $\pm$ 0.029\% (corresponding to an average absolute error of 0.400 $\pm$ 0.046 mK), an average median error of 0.223 $\pm$ 0.027\%, and an average maximum error of 1.09 $\pm$ 0.17\%. The best trial has a mean relative error of 0.21\% (corresponding to absolute error of 0.33 mK), a median error of 0.18\% and a maximum error of 0.94\%. Therefore, when trained and tested on the same data for the same number of trials, {\sc 21cmKAN} has a $1.1\times$ higher average error and $1.3\times$ higher maximum error than those reported for {\sc 21cmLSTM} (see Table~\ref{tab:results}), which currently is the most accurate emulator of the global 21 cm signal. Compared to {\sc 21cmVAE}, {\sc 21cmKAN} has a $1.4\times$ lower average error and $1.7\times$ lower maximum error.

When trained and tested on the physically-equivalent {\tt ARES} set for 20 trials, {\sc 21cmKAN} has an average relative mean error of 0.858 $\pm$ 0.075\% (corresponding to an average absolute error of 0.763 $\pm$ 0.061 mK), an average median error of 0.553 $\pm$ 0.046\%, and an average maximum error of 5.31 $\pm$ 1.51\%. The best trial has a mean relative error of 0.76\% (corresponding to 0.68 mK), a median error of 0.50\% and a maximum error of 4.58\%. The average emulation error of {\sc 21cmKAN} is therefore $\approx3\times$ larger when trained and tested on the {\tt ARES} set than on the {\sc 21cmGEM} set, and the maximum emulation error is $\approx5\times$ larger. The difference in {\sc 21cmKAN}'s emulation accuracy when trained on the {\sc 21cmGEM} set versus the {\tt ARES} set is likely due to differences in model parameterizations, parameter ranges, and sampling uniformity (see Section~\ref{subsec:model_comparison}).

\begin{table}
    \caption{Accuracy and Evaluation Speed Metrics of Global 21 cm Signal Emulators\label{tab:results}}
    \begin{tabular}{cccc}
    \toprule
    \addlinespace
    Emulator& Mean Error& Maximum Error & Speed\\
    &(\%)& (\%)& (ms)\\
    \addlinespace
    \hline
    {\sc 21cmKAN} & 0.25 & 1.09 & 3.7\\
    {\sc 21cmLSTM} & 0.22 & 0.82 & 46\\
    {\sc 21cmVAE} & 0.35 & 1.84 & 41.4\\
    {\tt globalemu} & 1.12 & 6.32 & 1.3\\
    {\sc 21cmGEM} & 1.59 & 10.55 & 160\\
    \bottomrule
    \end{tabular}
    \begin{tablenotes}
    \small
    \item The information provided for each emulator are the average mean and maximum rms errors across the full frequency range of $\approx$1700 {\sc 21cmGEM} test set signals (see Section~\ref{subsec:accuracy}), and the average per signal evaluation speed (see Section~\ref{subsec:speed}). The per signal prediction speed quoted for {\sc 21cmKAN} is the average speed from nested sampling analyses (see Section~\ref{subsubsec:posterioremulationresults}, Table~\ref{tab:multinest}), using the computational resources stated in Section~\ref{subsec:architecture}. The emulation errors and evaluation speeds quoted for {\sc 21cmLSTM} \citep{DorigoJones24}, {\sc 21cmVAE} \citep{21cmVAE}, {\tt globalemu} \citep{globalemu}, and {\sc 21cmGEM} \citep{Cohen20} are from their respective original papers that utilized different computational resources. The evaluation speeds of {\sc 21cmLSTM} and {\sc 21cmVAE} were measured using a GPU, while the evaluation speed of {\tt globalemu} was measured using a CPU (see the respective original papers for details).
    \end{tablenotes}
\end{table}

\subsection{Speed} \label{subsec:speed}
Two different speeds are relevant when assessing an emulator's efficiency and usefulness for parameter estimation: the training speed and the per sample evaluation speed. We report the training speed of {\sc 21cmKAN} as the average time to train one instance, and we report its evaluation speed as the average time to emulate one signal (i.e., predict $\delta T_b$ for all frequencies from a set of input parameters) during nested sampling analyses, including the time for data preprocessing and signal denormalization (see Section~\ref{subsec:data}; Equation~\ref{eqn:minmax}). Each reported speed of {\sc 21cmKAN} was measured using the same computational resources (see number and type of CPUs and GPU in Section~\ref{subsec:architecture}), which it naturally depends on.

Across 20 trials, the average time required to train {\sc 21cmKAN} on the {\sc 21cmGEM} set is only 10 minutes, which is very fast compared to existing emulators of the global 21 cm signal (see below). The average time required to train {\sc 21cmKAN} on the {\tt ARES} set (i.e., for 800 epochs rather than 400 epochs) is 19.2 minutes. When trained on the same set of {\sc 21cmGEM} signals and using the same computational resources, we find that {\sc 21cmKAN} trains $\approx75\times$ faster than {\sc 21cmLSTM}. {\sc 21cmKAN} trains significantly faster than {\sc 21cmLSTM} because of its fully parallel feedforward architecture as opposed to the sequential nature of LSTM cells, and because standard backpropagation is much simpler and less memory-intensive than the backpropagation through time required for recurrent NNs. We also find that {\sc 21cmKAN} trains $\approx3-4\times$ faster than the current emulator with the fastest evaluation speed, {\tt globalemu} \citep{globalemu}, when using the same computational resources. We trained the most recent version of {\tt globalemu} with all preprocessing steps turned on, using three hidden layers of 32 nodes each. {\sc 21cmKAN} converges more quickly during training because it learns expressive functional transformations that allow it to better model the global 21 cm signal shape.

Regarding the evaluation speed of {\sc 21cmKAN}, we find that the average per signal evaluation time (i.e., the total time of nested sampling analyses performed in Section~\ref{subsec:posterioremulation} divided by the total number of likelihood evaluations) is $3.7$ ms. When using {\sc 21cmKAN} and {\sc 21cmLSTM} to perform the same nested sampling analyses with the same computational resources, we find that {\sc 21cmKAN} evaluates $\approx5\times$ faster than {\sc 21cmLSTM}. Because here we do not perform nested sampling analyses using other emulators, we do not directly compare with their evaluation speeds and instead report in Table~\ref{tab:results} the speeds measured in their respective original papers that utilized different computational resources. Emulator training time is a more significant bottleneck in inference pipelines than evaluation speed, given the need to train an emulator with different parameterizations to constrain various combinations of physical parameters across multiple models. As it might be expected, the order of the evaluation times for {\sc 21cmLSTM}, {\sc 21cmKAN}, and {\tt globalemu} matches the order of the number of trainable parameters (i.e., weights and biases) in each emulator, with {\sc 21cmLSTM} having the highest and {\tt globalemu} the lowest; however, because of differences in network architectures and computational resources, the ratios of the emulator evaluation times do not directly match the ratios of their trainable parameters.

In summary, the extremely fast training and evaluation speed and low emulation error of {\sc 21cmKAN} altogether enable rapid and highly accurate Bayesian multi-parameter estimations of different parameterizations and models, which we carry out in Section~\ref{subsec:posterioremulation}. Efficient training is important for global 21 cm signal inference pipelines because emulator models with different parameterizations are needed to constrain the different cosmological and astrophysical parameters and their covariances across signal models. The speed and accuracy of {\sc 21cmKAN} highlight the effectiveness of KANs in learning physical mappings between relatively low-dimensional input and smooth output spaces that are traditionally computed using expensive PDE solvers.

\subsection{Posterior Emulation} \label{subsec:posterioremulation}
\begin{figure*}
    \includegraphics[scale=0.355]{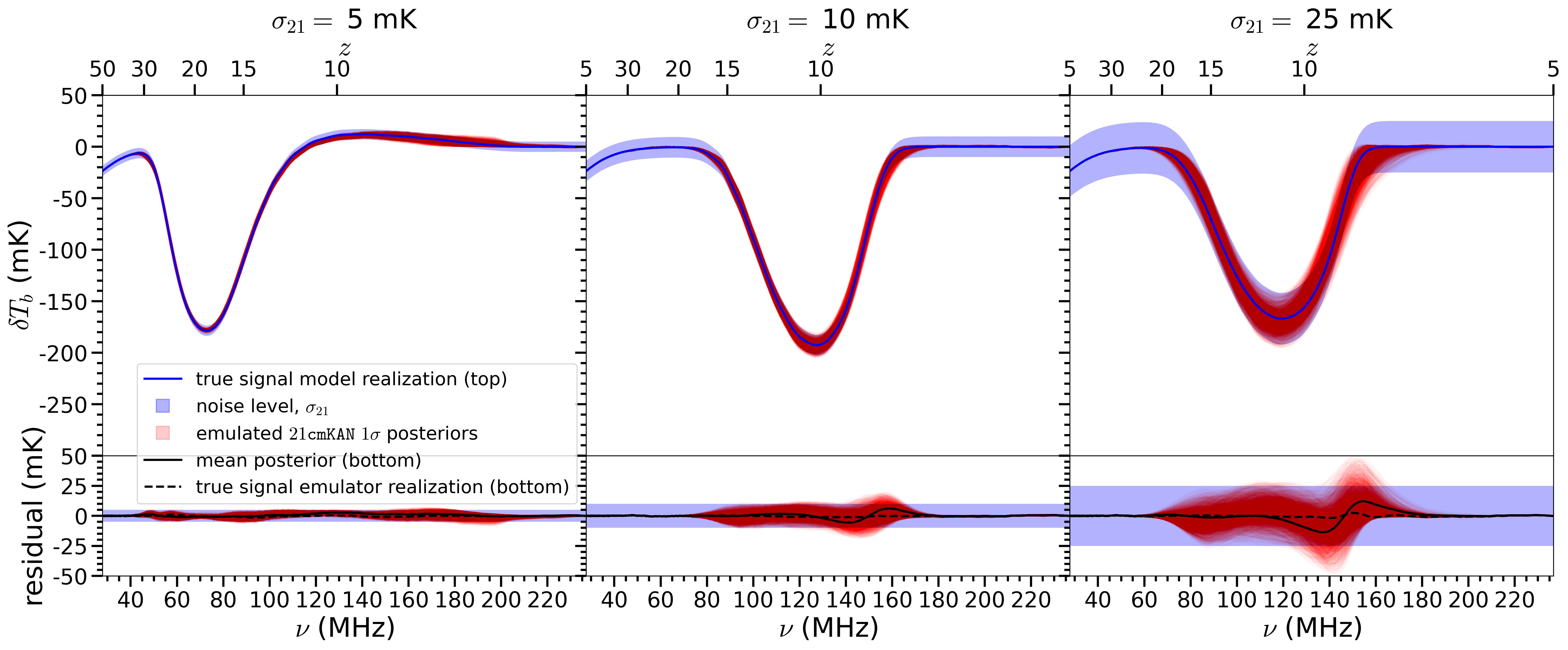}
    \caption{{\it Top:} Signal realizations of the $1\sigma$ posteriors (red, see Section~\ref{subsec:posterioremulation}) obtained from nested sampling analyses using {\sc 21cmKAN} to fit three global 21 cm signals (dark blue) randomly selected from the {\sc 21cmGEM} test set (see \citealt{DorigoJones24}) with added noise (light blue bands) of 5 mK (left), 10 mK (middle), and 25 mK (right). {\it Bottom:} Residuals between the corresponding true signal and each {\sc 21cmKAN} $1\sigma$ posterior (red, see Table~\ref{tab:multinest}), 
    the mean posterior (solid black), and the signal emulation (dashed black).\label{fig:posterior_21cmGEM}}
\end{figure*}

\begin{figure*}
    \includegraphics[scale=0.355]{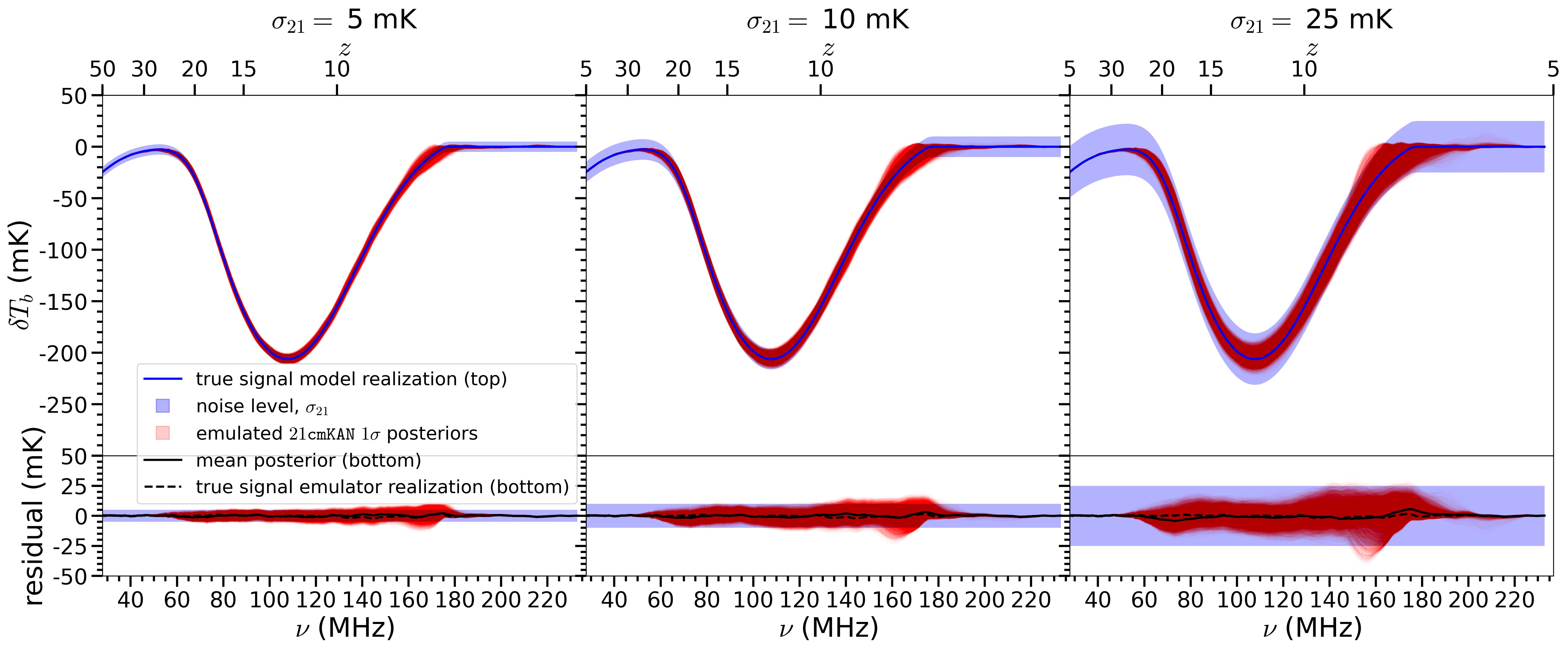}
    \caption{{\it Top:} Same as Figure~\ref{fig:posterior_21cmGEM} but when using {\sc 21cmKAN} to fit the same {\tt ARES} signal that was fit in \citet{DorigoJones23}.\label{fig:posterior_ARES}}
\end{figure*}

\subsubsection{Nested Sampling Analysis} \label{subsubsec:posterioremulationmethods}
We present results when using {\sc 21cmKAN}, in place of the full astrophysical model it is trained on, in the likelihood of Bayesian inference analyses to fit mock global 21 cm signals with added statistical noise and estimate its parameters. In doing so, we evaluate how well {\sc 21cmKAN} constrains synthetic 21 cm signals. As mentioned in Section~\ref{sec:intro}, because Monte Carlo sampling methods are computationally expensive, requiring $10^4$ to $10^6$ or more likelihood evaluations to explore multi-dimensional parameter spaces, emulators are desired to speed up or make feasible such analyses.~Our analyses here assume that systematic uncertainties have been properly accounted for, such as those from the beam-weighted foreground (see, e.g., \citealt{Bernardi16, Hibbard20, Anstey23, Sims23, Hibbard23, Pagano24, Saxena24}) and environment (see, e.g., \citealt{SARAS2, Kern20, Bassett21, Shen22, Murray22}).

We employ {\tt MultiNest} (\citealt{Feroz08, Feroz09, Feroz19}), which is a nested sampling Bayesian parameter inference tool (\citealt{Skilling04}; for reviews see \citealt{Ashton22}; \citealt{Buchner23}), to numerically sample posterior distributions, $P(\theta | \boldsymbol{D}, m)$, where $\theta$ are the parameters of the given physical model $m$, $\boldsymbol{D}$ is the mock data being fit, and $\pi$ are the parameter prior distributions that we assume to be uniform or log-uniform. Bayes' theorem defines the posterior as:
\begin{equation} \label{eqn:Bayes}
P(\theta | \boldsymbol{D}, m) = \frac{\mathcal{L} (\theta) \pi (\theta)}{{\it Z}},
\end{equation}
\noindent where $\mathcal{L}$ is the likelihood function and {\it Z} is the Bayesian evidence, which can be used for model comparison. The log-likelihood function we sample from is multivariate and assumes a Gaussian probability density function (PDF) for the residuals: $\log \mathcal{L}(\theta) \propto [\boldsymbol{D} - m(\theta)]^T \boldsymbol{C}^{-1} [\boldsymbol{D} - m(\theta)]$, where $\boldsymbol{C}$ is the (diagonal) noise covariance array containing the square of the noise estimate $\sigma_{21}$. Nested sampling methods iteratively remove prior volume regions with lower likelihood and simultaneously compute the evidence and posterior (see e.g., \citealt{Lemos23} for an in-depth description of nested sampling algorithms). Compared to MCMC, nested sampling better constrains complex, multimodal posterior distributions \citep{Buchner23}, which have been found when fitting 21 cm data (e.g., \citealt{Bevins22b, DorigoJones23, Breitman24}; also see \citealt{Saxena24}). For the {\tt MultiNest} sampling parameter values, we use the default sampling efficiency (i.e., 0.8), evidence tolerance of 0.1, and 1200 initial ``live'' points (i.e., $3\times$ the default), which we find result in consistent, converged posteriors.

We fit the same three mock 21 cm signals that were randomly selected from the {\sc 21cmGEM} test set and fit using {\sc 21cmLSTM} in \citet{DorigoJones24}, and we also fit the same {\tt ARES} signal that was fit using {\tt globalemu} in \citet{DorigoJones23}, to allow for direct comparisons to the nested sampling results from these works. The statistical noise added to each signal is Gaussian-distributed, white noise, ranging from $\sigma_{21}=5$ mK or 10 mK (referred to as ``optimistic''; see \citealt{REACH}) to $\sigma_{21}=25$ mK or 50 mK (referred to as ``standard''). We note that $\sigma_{21}=$ 5, 10, 25, and 50 mK correspond to integration times of about 7100, 1800, 300, and 70 hours, respectively, when assuming $\Delta\nu=0.5$ MHz in the ideal radiometer sensitivity equation (e.g., \citealt{Kraus66}) and $\nu=30$ MHz to compute the antenna temperature from galactic synchrotron radiation \citep{Haslam}.

To fit the {\sc 21cmGEM} signals, we used an instance of {\sc 21cmKAN} trained on the {\sc 21cmGEM} set that has test set mean rms error of 0.26\% (corresponding to 0.42 mK) and maximum error of 1.27\%. To fit the {\tt ARES} signal, we used an instance of {\sc 21cmKAN} trained on {\tt ARES} that has test set mean rms error of 0.81\% (corresponding to 0.72 mK) and maximum error of 6.41\%. Each trained network is consistent with the respective average accuracy found in Section~\ref{subsec:accuracy}, and all fits used the same computational resources (see Section~\ref{subsec:architecture}).

\subsubsection{Posterior Results} \label{subsubsec:posterioremulationresults}
\begin{table}
    \centering
    \caption{Summary of Nested Sampling Analyses\label{tab:multinest}}
    \begin{tabular}{ccccccc}
    \toprule
    \addlinespace
        Signal Fit&$\sigma_{21}$&$n_{\rm eval}$&$f_{\rm acc}$&$\log {\it Z}$&$t_{\rm eval}$&$\sigma_{\rm post}$\\
        &(mK)& & & &(ms)&(mK)\\
    \addlinespace
    \hline
    \multirow{3}{*}{\sc 21cmGEM} &5&136,397&0.190&-256.6&3.7&1.7\\
    &10&64,749&0.338&-253.4&4.2&2.9\\
    &25&47,711&0.363&-250.9&4.4&7.8\\
    \cline{1-7}
    \multirow{4}{*}{\tt ARES} &5&335,760&0.090&-258.6&3.2&1.9\\
    &10&208,161&0.126&-255.0&3.4&3.2\\
    &25&132,493&0.157&-250.8&3.5&5.7\\
    &50&82,675&0.208&-248.1&3.8&--\\
    \bottomrule
    \end{tabular}
    \begin{tablenotes}
    \small
    \item The information provided for each fit are the noise level of the mock 21 cm signal ($\sigma_{21}$), the total number of likelihood evaluations ($n_{\rm eval}$), the final acceptance rate ($f_{\rm acc}$), the final evidence ($\log {\it Z}$), the time per likelihood evaluation ($t_{\rm eval}$), and the $1\sigma$ rms error of all posterior samples ($\sigma_{\rm post}$). Each $\log {\it Z}$ has error of $\pm$0.1. The nested sampling analysis and results are described in Section~\ref{subsec:posterioremulation}. The $1\sigma$ posterior signal realizations and residuals with respect to the true {\sc 21cmGEM} and {\tt ARES} signals are shown in Figures~\ref{fig:posterior_21cmGEM} and~\ref{fig:posterior_ARES}, respectively. The full posterior distributions for the {\sc 21cmGEM} fits with $\sigma_{21}=$ 25 mK and $\sigma_{21}=$ 5 mK are shown in Figures~\ref{fig:appendix_corner_25mK} and~\ref{fig:appendix_corner_5mK}, respectively, and for the {\tt ARES} fit with $\sigma_{21}=$ 50 mK in Figure~\ref{fig:appendix_corner_50mK}.
    \end{tablenotes}
\end{table}

In the top panels of Figures~\ref{fig:posterior_21cmGEM} and~\ref{fig:posterior_ARES}, we present sets of posterior signal realizations (shown in red), when using {\sc 21cmKAN} to fit {\sc 21cmGEM} and {\tt ARES} mock 21 cm signals (shown in dark blue), respectively, with added noise levels (shown in light blue) of $\sigma_{21}=$ 5, 10, and 25 mK, from left to right. The $1\sigma$ emulated posteriors are shown in red, which we define as the 68\% of samples with the lowest relative rms error with respect to the true signal (Equation~\ref{eqn:rel_error}), although we note that the distribution of rms errors is not exactly Gaussian. The bottom panels show the residuals between the true signals and the $1\sigma$ posteriors (red), the mean posterior (solid black), and the emulator realization of the true signal (i.e., $m_{{\tt 21cmKAN}}(\theta_0)$; dashed black). Table~\ref{tab:multinest} summarizes each fit, and the marginalized 1D and 2D posterior distributions for three of the fits are presented in Appendix~\ref{sec:full_posteriors} along with the physical parameter values, $\theta_0$, used to generate each mock signal. We note that the fits using {\sc 21cmKAN} to constrain the {\sc 21cmGEM} signals completed in 3 to 8 minutes, and the fits using {\sc 21cmKAN} to constrain the {\tt ARES} signal completed in 5 to 18 minutes, depending on the noise level $\sigma_{21}$ (see Table~\ref{tab:multinest}).

As expected, the fits obtained using {\sc 21cmKAN} consistently improve for decreasing $\sigma_{21}$, following the increase in constraining power for longer assumed integration times, and this does not depend on the particular mock signal being fit. As seen visually in the bottom panels of Figures~\ref{fig:posterior_21cmGEM} and~\ref{fig:posterior_ARES} and quantitatively in Table~\ref{tab:multinest}, we find that the mean and $1\sigma$ rms errors of the posteriors are consistently $\approx3\times$ less than $\sigma_{21}$ and approach the emulator error (see Section~\ref{subsubsec:posterioremulationmethods}) as $\sigma_{21}$ decreases, which facilitates unbiased parameter posterior distributions (see below). For the {\sc 21cmGEM} signal fits, the mean relative rms error (Equation~\ref{eqn:rel_error}) between all the emulated posteriors and the true signal is 0.91\% (corresponding to 1.63 mK absolute error) for the fit done with $\sigma_{21}=5$ mK, 1.42\% (2.75 mK) for $\sigma_{21}=10$ mK, and 4.15\% (6.94 mK) for $\sigma_{21}=25$ mK. For the {\tt ARES} signal fits, the posterior mean relative rms error is 0.79\% (corresponding to 1.62 mK absolute error) for $\sigma_{21}=5$ mK, 1.38\% (2.85 mK) for $\sigma_{21}=10$ mK, and 2.93\% (6.05 mK) for $\sigma_{21}=25$ mK. We note that the {\tt ARES} signal fits required $\approx3\times$ more likelihood evaluations to converge compared to the {\sc 21cmGEM} signal fits with the same $\sigma_{21}$ (see Table~\ref{tab:multinest}), which reflects the $\approx3\times$ higher average emulation error for {\sc 21cmKAN} when trained on the {\tt ARES} set than when trained on the {\sc 21cmGEM} set (Sections~\ref{subsec:accuracy} and~\ref{subsec:model_comparison}).

In general, the {\sc 21cmGEM} signal fits using {\sc 21cmKAN} are extremely similar to the same fits performed using {\sc 21cmLSTM} presented in \citet{DorigoJones24}. Furthermore, as described below, the {\tt ARES} signal fits using {\sc 21cmKAN} are slightly better than the same fits performed using {\tt globalemu} presented in \citet{DorigoJones23}. These results are expected, given that the average emulation error of {\sc 21cmKAN} is only slightly higher than that of {\sc 21cmLSTM} and is significantly lower than that of {\tt globalemu} (see Section~\ref{subsec:accuracy}; Table~\ref{tab:results}). For the {\sc 21cmGEM} signal fit with $\sigma_{21}=25$ mK (Figure~\ref{fig:appendix_corner_25mK}), {\sc 21cmKAN} obtains well-constrained and unbiased posteriors (i.e., captures the true parameter values to within $2\sigma$) for $f_*$, $V_c$, and $\tau$, and for the $\sigma_{21}=5$ mK fit (Figure~\ref{fig:appendix_corner_5mK}) the constraints become unbiased for $f_X$ and $\nu_{\rm min}$. The same fits performed using {\sc 21cmLSTM} in \citet{DorigoJones24} obtained nearly equivalent final evidences and $1\sigma$ posterior signal residuals (see their Table 3) and very similar but slightly better posterior distributions (see their Figures B1 and B2). For the {\tt ARES} signal fit with $\sigma_{21}=50$ mK (Figure~\ref{fig:appendix_corner_50mK}), {\sc 21cmKAN} obtains unbiased posteriors for $c_X$, $f_{\rm esc}$, $T_{\rm min}$, and $f_{\rm \star,0}$, capturing their true values within $2\sigma$. The posteriors obtained using {\tt globalemu} in \citet{DorigoJones23} were similar but slightly less centered on the true parameter values and fewer 2D posteriors captured the true values within $2\sigma$ (see their Figure 6), which is reflected in the final evidence being slightly worse (see their Table 2). These mock 21 cm posterior constraints support studies that have found it necessary to combine complementary summary statistics or data sets to break degeneracies between some physical parameters (e.g., \citealt{Qin20, Chatterjee21, DorigoJones23, Breitman24}).

To summarize, {\sc 21cmKAN} obtains unbiased physical parameter posteriors when used in Bayesian inference analyses to fit mock global 21 cm signals from two different simulations -- {\sc 21cmGEM} and {\tt ARES} -- with standard or extremely optimistic noise levels. {\sc 21cmKAN} has signal and posterior emulation accuracy on par with the most accurate emulator, {\sc 21cmLSTM}, and requires only 15 to 38 minutes for the entire process from beginning of training to completing the multidimensional nested sampling analyses presented here, when using a typical GPU (see resources in Section~\ref{subsec:architecture}). The combination of exceptionally high speed and accuracy makes {\sc 21cmKAN} valuable for constraining real data of the 21 cm signal because different physical model parameterizations will need to be tested.

\subsection{Visualizing and Interpreting Learned {\sc 21cmKAN} Activation Functions When Trained on Different Models} \label{subsec:model_comparison}
\begin{figure*}
    \includegraphics[scale=0.255]{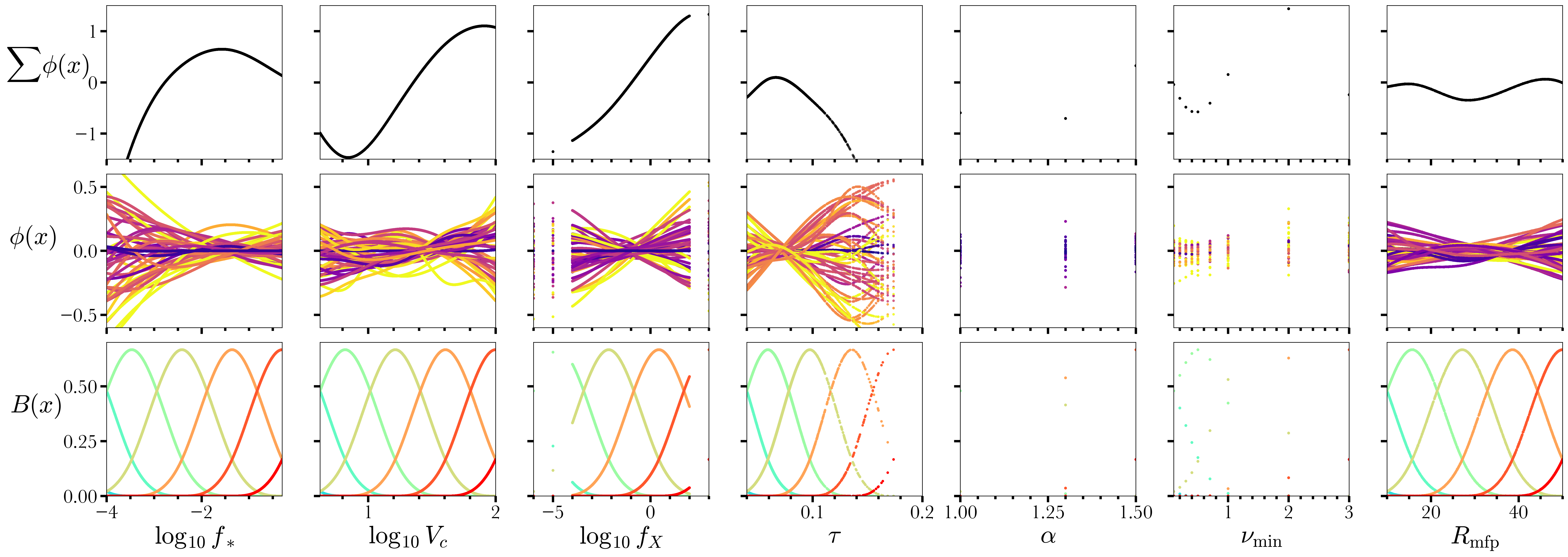}
    \caption{Univariate, nonlinear activation functions in the first hidden layer for each input physical parameter, learned by a trial of {\sc 21cmKAN} trained on the {\sc 21cmGEM} set and evaluated at the parameter values of the 24,562 training set signals (see Section~\ref{subsec:model_comparison}). {\it Top:} Aggregated activation function output that is the input for the second hidden layer nodes. {\it Middle:} B-spline activation functions for all 44 edges, with color depicting the largest basis function coefficient value for each (see Equation~\ref{eqn:activation}). Yellow activations have larger coefficients of their constituent basis functions, while purple activations have smaller, or less influential, basis functions. {\it Bottom:} Local B-spline basis functions.\label{fig:splines_GEM}}
\end{figure*}

\begin{figure*}
    \includegraphics[scale=0.22]{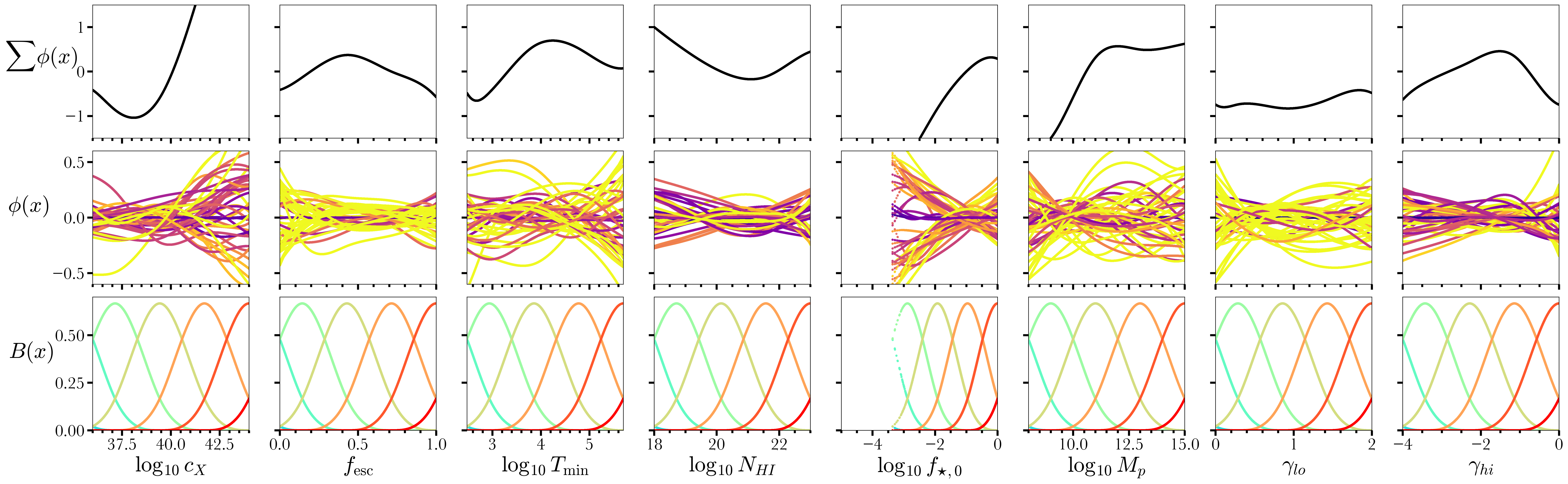}
    \caption{Same as Figure~\ref{fig:splines_GEM} but for {\sc 21cmKAN} trained on the {\tt ARES} set and evaluated at the parameter values of the 23,896 training set signals. The y-axis ranges are the same as in Figure~\ref{fig:splines_GEM} to show that the per-edge activations are generally higher magnitude when trained on the {\tt ARES} set.\label{fig:splines_ARES}}
\end{figure*}

We now describe the {\sc 21cmKAN} architecture in more detail and present visualizations of the learned activation functions, which help interpret the emulator’s behavior and the sensitivity of the 21 cm signal to different physical parameters. For a general overview of the KAN and how it compares to other types of NNs, see Section~\ref{subsec:KAN} and Figures~\ref{fig:KAN} and~\ref{fig:full}. By simply observing the components of the trained emulator, we demonstrate that {\sc 21cmKAN} conveniently provides an intuitive understanding of its predictions that is complementary to the information gained from more rigorous techniques like saliency mapping, feature importance or attribution, and activation maximization \citep{Montavon18, Lipton18}. The interpretability of {\sc 21cmKAN} further validates its performance and motivates transferability to other models and tasks beyond the emulation results presented in Sections~\ref{subsec:accuracy} and~\ref{subsec:posterioremulation} (e.g., \citealt{KAN, Belfiore25}).

Let a given hidden layer have input dimensionality $p$ and output dimensionality $q$, such that the default {\sc 21cmKAN} trained on the {\sc 21cmGEM} set has $p=7$ and $q=44$ for the first hidden layer. Each edge connecting input node $p$ to output node $q$ learns a univariate activation function, $\phi_{q,p}(x_p)$, and applies it to the corresponding input parameter $x_p$, such that there are $pq$ different activations per layer. Each per-edge function, hereafter denoted $\phi(x)$, is a B-spline curve \citep{spline} that is the weighted sum of local basis functions $B_i(x)$ with learnable coefficients $c_i$ (see \citealt{KAN}):
\begin{equation} \label{eqn:activation}
\phi(x)\approx\sum c_iB_i(x).
\end{equation}
\noindent For a B-spline parameterized by $G$ grid intervals (i.e., $G+1$ grid points) and individual splines of order $k$, there are $G+k$ basis functions, which sum to one at each $x_p$, $\sum B_i(x_p) = 1$, making KANs a form of nonlinear dimensionality reduction or feature extraction (see also multivariate adaptive regression splines; \citealt{Friedman91, BMARS, BASS}). At the start of training, the basis function coefficients and thus B-splines are zero, and the splines become nonlinear and grow in magnitude as the $c_i$ are learned. The input to each node in a subsequent layer is the sum of all the incoming post-activations across $p$, $\sum\phi(x)$ (see also multiplicative KANs; \citealt{MultKAN}). In summary, KANs learn and apply flexible nonlinear activation functions on the edges and sum them at the nodes to map input to output, whereas traditional MLPs learn scalar weights on the edges and apply fixed, pre-determined activations at the nodes.

To visualize the KAN activations, we present in Figures~\ref{fig:splines_GEM} and~\ref{fig:splines_ARES} the basis functions ($B(x)$, bottom panels), per-edge B-spline activations ($\phi(x)$, middle panels), and their node-wise sums ($\sum\phi(x)$, top panels) in the first hidden layer for each physical parameter that were learned by {\sc 21cmKAN} when trained on the {\sc 21cmGEM} and {\tt ARES} data sets, respectively. The first hidden layer is more interpretable or explainable than subsequent layers \citep{Montavon18, Lipton18} because the learned activations depict how each input physical parameter contributes to the output. To generate these plots, we evaluated the same trials of {\sc 21cmKAN} used for posterior emulation in Section~\ref{subsec:posterioremulation} at the parameter values of the respective training set (see ranges in Table~\ref{tab:params}).

The KAN activation functions for almost all physical parameters are nonlinear and vary significantly across the parameters and edges, indicating that {\sc 21cmKAN} learns distinct contributions of each parameter to the output signal $\delta T_b$. The learned activations and basis function coefficients can change between different trials of {\sc 21cmKAN}, which is expected because NNs often converge to different local minima, but this stochasticity may preclude direct connections between the activation functions and relevant physical equations. As seen in the middle panels of Figures~\ref{fig:splines_GEM} and~\ref{fig:splines_ARES}, larger magnitude activations correlate with larger basis function coefficients (yellow), whereas smaller activations generally correspond to smaller $c_i$ (purple), as expected from Equation~\ref{eqn:activation}. The peaks of the activations typically coincide with the local basis function that has the largest learned coefficient, highlighting physical parameter values with greater impact on the output 21 cm signal. {\sc 21cmKAN} learns near-zero coefficients for the first three basis functions, and so the first three $B(x)$ are zero across all $x_p$ and parameters.

We find that physical parameters with larger magnitude per-edge activations learned by {\sc 21cmKAN} correspond to those with greater influence on the 21 cm signal prediction. This relationship between activation magnitude and feature importance was also found in \citet{KAN} as an example of KANs being inherently interpretable and complementary to \textit{post-hoc} explainable machine learning approaches (see their Section 4.3) and could be utilized for domain adaptation. Additionally, activations with greater variation along parameter values mean that small changes in the parameter meaningfully affect the output signal and imply the signal is strongly sensitive to that parameter, which is also indicated by more constrained posteriors (see Section~\ref{subsubsec:posterioremulationresults}). The learned activations across all 20 trials for {\sc 21cmGEM} parameters $f_{\star}$, $V_c$, and $f_X$ and {\tt ARES} parameters $c_X$ and $T_{\rm min}$ are larger in magnitude and present larger variations, and their posterior distributions are correspondingly well constrained in Figures~\ref{fig:appendix_corner_5mK} and~\ref{fig:appendix_corner_50mK}, respectively. In contrast, the {\sc 21cmGEM} parameter $R_{\rm mfp}$ as well as the {\tt ARES} parameters $\log N_{\rm H \RomanNumeralCaps{1}}$ and $\gamma_{\rm hi}$ have activations with less variation and magnitude, and they also have the least constrained posteriors. These results are consistent with previous analyses (see, e.g., Figure 4 of \citealt{21cmVAE}) that find $R_{\rm mfp}$ to have the smallest impact on the {\sc 21cmGEM} signals and $f_{\star}$, $V_c$, and $f_X$ to have the largest impact.

Figures~\ref{fig:splines_GEM} and~\ref{fig:splines_ARES} also show that the sampling in the training sets is less uniform for certain parameters than others. For {\sc 21cmGEM}, $\alpha$ and $\nu_{\rm min}$ are sparsely sampled, and the lower end of $f_X$ and upper end of $\tau$ are less uniformly sampled, while the other parameters are uniformly sampled. For {\tt ARES}, the lower end of $f_{\rm \star,0}$ is less uniformly sampled, while all of the other parameters are uniformly sampled. As explained in Section 2 of \citet{Cohen20}, in creating the {\sc 21cmGEM} set, a few values were chosen for $\alpha$ and $\nu_{\rm min}$ to construct the spectral energy distribution (SED) power-laws of high-$z$ X-ray sources. The {\tt ARES} model treats X-ray SEDs in a fundamentally different manner than the {\sc 21cmGEM} model, as explained in \citet{Mirocha14} and \citet{Mirocha17}, by explicitly solving the cosmological radiative transfer equation rather than scaling the X-ray heating efficiency by $f_X$. The sparse sampling of $\alpha$ and $\nu_{\rm min}$ hinders the variation in the {\sc 21cmGEM} training set and restricts {\sc 21cmKAN} to learn from fewer parameter values, which effectively simplifies the mapping. Combined with the fact that the {\sc 21cmGEM} set contains one fewer physical parameter than the {\tt ARES} set, we infer that the non-uniform sampling of $\alpha$ and $\nu_{\rm min}$ contributes to the $\approx3\times$ lower average emulation error of {\sc 21cmKAN} and $2\times$ faster time for it to reach a stable validation loss when trained on the {\sc 21cmGEM} set than the {\tt ARES} set (see Section~\ref{subsec:accuracy}; Figure~\ref{fig:appendix_loss}). This interpretation is consistent with the principal component analysis decompositions performed in \citet{DorigoJones24}, which identified lower statistical variation among the signals in the {\sc 21cmGEM} set than those in the {\tt ARES} set. We leave for future work a detailed comparison of the {\sc 21cmGEM} and {\tt ARES} models and a potential domain adaptation between them enabled by {\sc 21cmKAN}.

\section{Conclusions} \label{sec:conclusions}
In this paper, we presented {\sc 21cmKAN}, which is a publicly-available\footnoteref{code} Kolmogorov-Arnold Network-based emulator of the global 21 cm signal. KANs\footnoteref{KAN} \citep{KAN} provide a fundamental shift in the architecture of fully-connected NNs, by explicitly learning the nonlinear transformations, or activation functions, that best model complex relationships in data (Figures~\ref{fig:KAN} and~\ref{fig:full}). The expressivity of KANs makes them an improvement in accuracy and speed over traditional fully-connected NNs, which use fixed activation functions, when modeling certain structured, lower-dimensional functions or PDEs often found in science (e.g., \citealt{Shukla24, Cui25, CuiShuangwei25, Shi25}). {\sc 21cmKAN} has low emulation error similar to the most accurate existing emulator of the global 21 cm signal, {\sc 21cmLSTM} \citep{DorigoJones24} and trains $\approx75\times$ faster than {\sc 21cmLSTM} (Figure~\ref{fig:err_hist}; Table~\ref{tab:results}; Section~\ref{subsec:accuracy}; Section~\ref{subsec:speed}).

The combination of low emulation error and extremely fast training and evaluation speed of {\sc 21cmKAN} enables rapid and highly accurate Bayesian inferences of multiple physical models. Emulators will need to be trained for various cosmological and astrophysical models of the global 21 cm signal to constrain their parameters. Posterior distributions generally change with the addition or removal of parameters or data sets based on their covariances and constraints (e.g., \citealt{Qin20, Chatterjee21, DorigoJones23, Breitman24}), and so it will be important to explore different combinations of parameters to robustly fit and fully exploit measurements of the signal. The efficiency of {\sc 21cmKAN} eliminates emulator training as a bottleneck for such comprehensive studies using inference pipelines. We trained and tested {\sc 21cmKAN} on two popular physical models of the global 21 cm signal and employed {\sc 21cmKAN} in Bayesian parameter estimation analyses to fit synthetic data with varying levels of added observational noise (Section~\ref{subsec:posterioremulation}). {\sc 21cmKAN} obtained signal fits with rms errors $\approx3\times$ less than the assumed noise (Figures~\ref{fig:posterior_21cmGEM},~\ref{fig:posterior_ARES}; Table~\ref{tab:multinest}) as well as unbiased posterior distributions for multiple well-constrained parameters in the two models (Figures~\ref{fig:appendix_corner_5mK},~\ref{fig:appendix_corner_50mK}), which are consistent with the results when using {\sc 21cmLSTM} to fit the same signals \citep{DorigoJones24}. {\sc 21cmKAN} evaluated each signal prediction in just 3.7 ms on average, when utilizing a typical A100 GPU, which enabled it to perform end-to-end training and posterior emulation altogether in under 30 minutes.

We found that the transparent nature of the {\sc 21cmKAN} architecture allows the user to conveniently infer the influence of different physical parameters on the global 21 cm signal by simply inspecting the learned activation functions (Section~\ref{subsec:model_comparison}; Figures~\ref{fig:splines_GEM},~\ref{fig:splines_ARES}). Physical parameters with larger magnitude and highly variable activation functions also have tightly constrained posterior distributions, which altogether imply high sensitivity of the 21 cm signal to these parameters. In contrast, physical parameters with lower magnitude or less variable learned activations are also those with poorly constrained posterior distributions, both of which imply that the 21 cm signal is weakly sensitive to these parameters. Such intuitive understanding of the emulator's innerworkings and predictions complement standard explainable machine learning tests and build further trust in the model \citep{Montavon18, Lipton18, KAN}. Beyond this, the simple and flexible architecture of {\sc 21cmKAN} could enable mapping between different models of the global 21 cm signal to unify their physical assumptions and thus improve the robustness and generalization of signal modeling in inference pipelines.

{\sc 21cmKAN} demonstrates the effectiveness and expressive nature of learnable neural activations for modeling the global 21 cm signal. While this work focused on a specific set of models and parameterizations, {\sc 21cmKAN} can be readily trained on other models or adapted to approximate a range of functions that may benefit from its unique architecture. More broadly, KANs can complement existing recurrent or convolutional NNs \citep{Genet24, Cheon24} and may prove to be valuable for other astrophysical or cosmological applications where deep learning has already shown success (e.g., \citealt{Liu19, Bahauddin21, PirasSpurio23, Breitman24, Kamai25}).

\begin{acknowledgments}
We thank the anonymous reviewer and Jordan Mirocha for their valuable feedback. This work utilized the Alpine high performance computing (HPC) cluster resource at the University of Colorado Boulder. Alpine is jointly funded by the University of Colorado Boulder, the University of Colorado Anschutz, Colorado State University, and the National Science Foundation (award \#2201538 and \#2322260). We acknowledge support by NASA APRA grant award 80NSSC23K0013, a subcontract from UC Berkeley (NASA award 80MSFC23CA015) to the University of Colorado (subcontract \#00011385) for science investigations involving the LuSEE-Night lunar far side mission, and NASA award 80NSSC22K1264 to support radio astrophysics from the Moon.
\end{acknowledgments}

\software{{\sc numpy} \citep{numpy}, {\sc matplotlib} \citep{matplotlib}, {\sc scipy} \citep{scipy}, {\sc PyTorch} \citep{pytorch}, KAN\footnoteref{KAN} \citep{KAN}, Ef-KAN\footnoteref{EfKAN}, {\sc jupyter} \citep{jupyter}, {\tt MultiNest} \citep{Feroz09}.}

\appendix
\restartappendixnumbering
\section{Emulation error for different {\sc 21cmKAN} architectures} \label{sec:diffarchs}
\begin{figure}[t]
    \includegraphics[width=\textwidth]{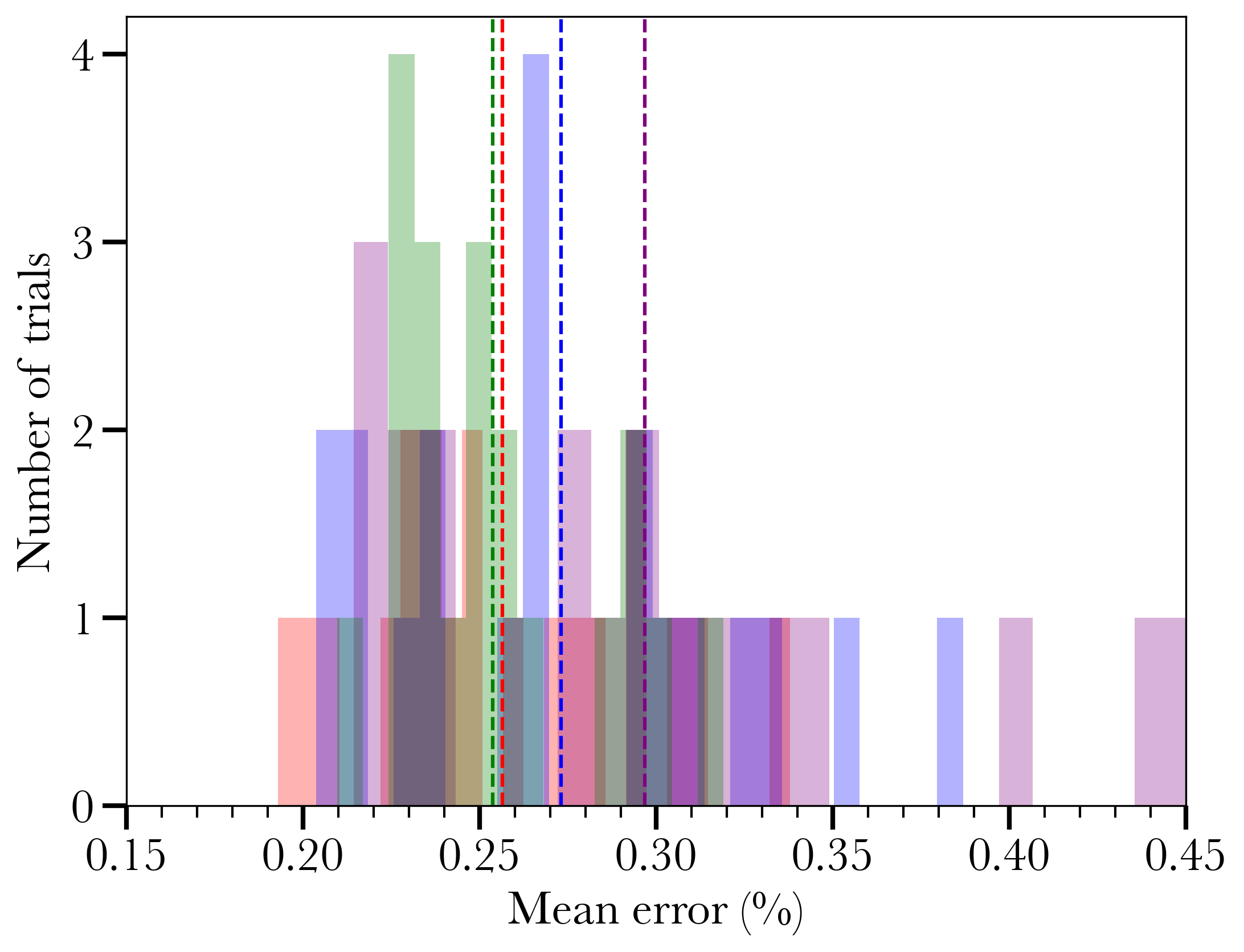}
    \caption{Emulation error (Equation~\ref{eqn:rel_error}) of {\sc 21cmKAN} using four different architectures when trained and tested on the {\sc 21cmGEM} set for 20 trials each. The average error for each architecture is indicated by the respective dashed line. The architecture hyperparameters varied are the numbers of nodes ($N$) per hidden layer (i.e., [$N_1$, $N_2$, $N_3$]), the B-spline grid size ($G$), the spline order ($k$), and the batch size ($b$). The green histogram used the default architecture (see Section~\ref{subsec:architecture}) of [$N_1$, $N_2$, $N_3$]$=$[44, 44, 71], $G=7$, $k=3$, and $b=100$. The red histogram used [$N_1$, $N_2$, $N_3$]$=$[20, 45, 74], $G=11$, $k=6$, $b=119$, the blue histogram [$N_1$, $N_2$, $N_3$]$=$[55, 63, 71], $G=11$, $k=7$, $b=84$, and the purple histogram [$N_1$, $N_2$, $N_3$]$=$[45, 95, 100], $G=11$, $k=5$, $b=195$.\label{fig:appendix_diffarchs}}
\end{figure}

In Figure~\ref{fig:appendix_diffarchs}, we present the emulation error of {\sc 21cmKAN} when trained and tested on the {\sc 21cmGEM} set using four different architectures, each with 20 trials performed (see caption for details). For the green, red, blue, and purple histograms, the average training time using each architecture is 10.1, 13.6, 20.2, and 8.0 minutes, respectively. We decided to use the green architecture as the default for {\sc 21cmKAN} because it results in the lowest average emulation error while still maintaining fast training time because of its relatively large batch size and low number of B-spline basis functions (see Section~\ref{subsec:model_comparison}).

\restartappendixnumbering
\section{Loss Curves for Validation and Training Sets} \label{sec:loss}
\begin{figure*}[t]
    \includegraphics[width=\textwidth]{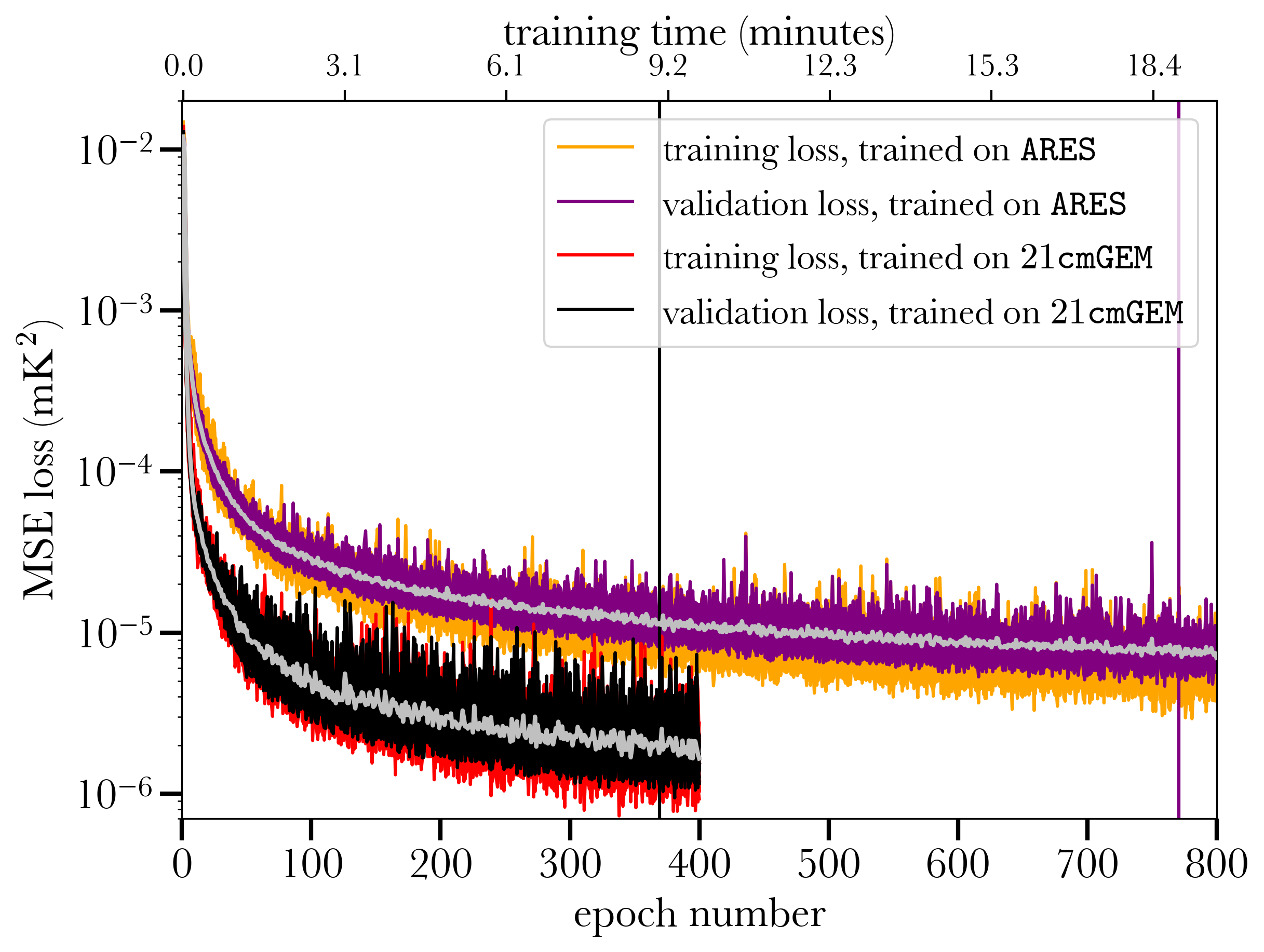}
    \caption{Loss versus training epoch number for validation and training sets for 20 trials of {\sc 21cmKAN} trained on the {\sc 21cmGEM} (black and red, respectively) and {\tt ARES} (purple and orange, respectively) sets (see Section~\ref{sec:methods}). The top axis shows the approximate training time at each epoch. The vertical lines indicate the average epochs of the minimum validation losses, which are 369 for {\sc 21cmGEM} and 771 for {\tt ARES}. Note that {\sc 21cmKAN} must be trained for $2\times$ longer on the {\tt ARES} set to reach a stable mean validation loss (silver curves; see Section~\ref{subsec:model_comparison}).\label{fig:appendix_loss}}
\end{figure*}

Figure~\ref{fig:appendix_loss} shows the distribution of MSE loss (Equation~\ref{eqn:mse}) for the training and validation sets for 20 identical trials of the default {\sc 21cmKAN} trained on the {\sc 21cmGEM} data and also when trained on the {\tt ARES} data (see Section~\ref{subsec:architecture}). The mean validation losses (shown by silver curves) reach stable values at the final training epochs, which indicates sufficient training and that the network is still able to generalize to unseen signals and is not overfitting the training set. We note that we trained {\sc 21cmKAN} on each set for more epochs and the average test set emulation errors did not significantly improve, which is also seen by the fact that the minimum validation loss does not occur at the end of the training on average (vertical lines).

\restartappendixnumbering
\section{Posterior distributions from fitting mock global 21 cm signals} \label{sec:full_posteriors}
\begin{figure*}[t]
    \includegraphics[width=\textwidth]{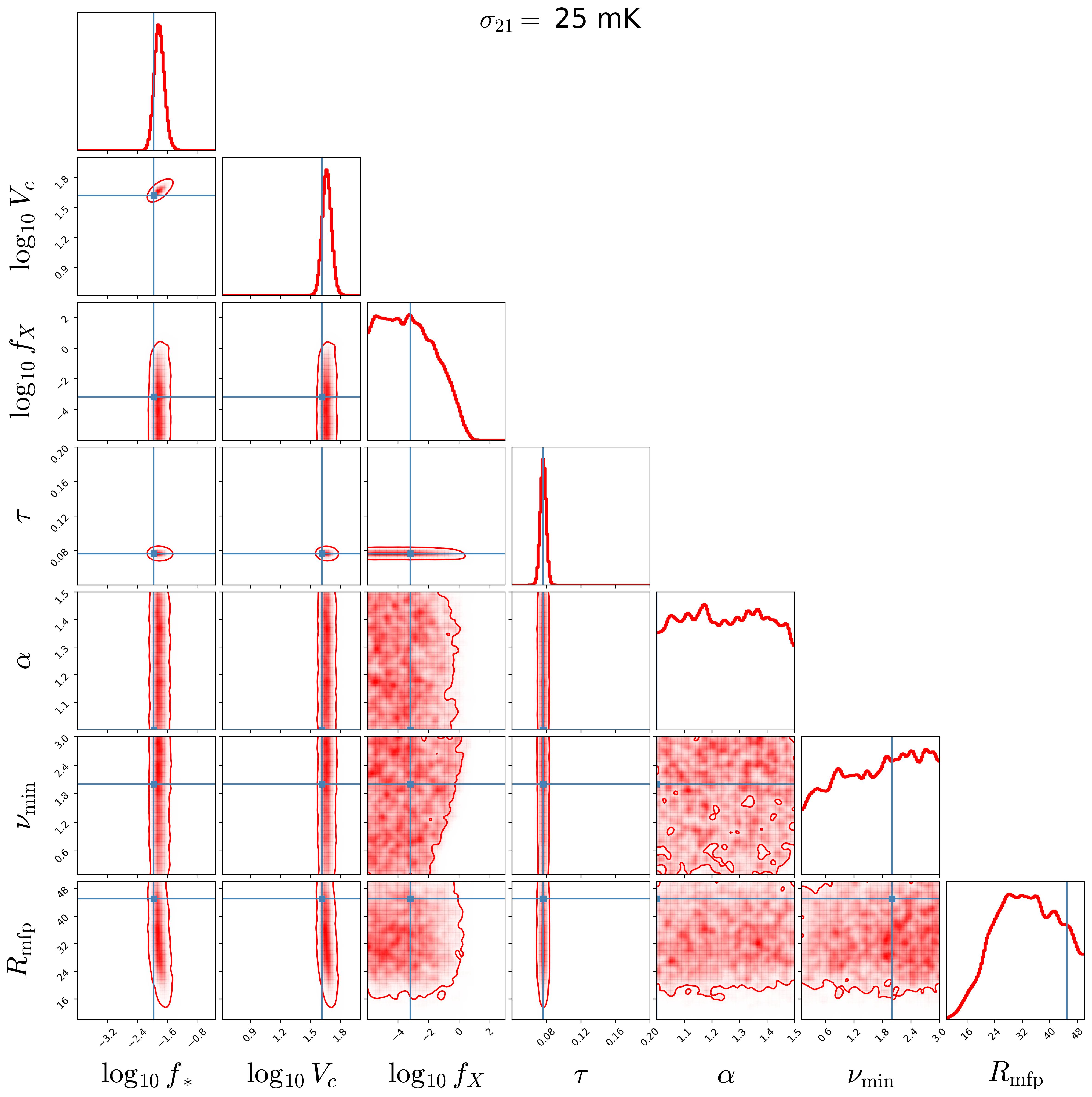}
    \caption{Marginalized 1D and 2D posteriors for seven astrophysical parameters obtained when using {\sc 21cmKAN} to fit a mock global 21 cm signal from the {\sc 21cmGEM} test set with the standard observational noise of $\sigma_{21}=$ 25 mK (see the right panel of Figure~\ref{fig:posterior_21cmGEM}; Section~\ref{subsec:posterioremulation}). These parameters control the SFE, emission of UV and X-ray photons, and CMB optical depth (see Table~\ref{tab:params}). Blue vertical and horizontal lines indicate the parameter values used to generate the mock signal being fit: $\theta_0=$($f_*$, $V_c$, $f_X$, $\tau$, $\alpha$, $\nu_{\rm min}$, $R_{\rm mfp}$)=(1.1020955$\times10^{-2}$, 41.533981, 6.4696016$\times10^{-4}$, 7.6195940$\times10^{-2}$, 1.0, 2.0, 45.0). Contour lines in the 2D posteriors represent the 95\% confidence levels, and sample density colormaps are shown. Axis ranges are the full training set prior ranges given in Table~\ref{tab:params}.\label{fig:appendix_corner_25mK}}
\end{figure*}

\begin{figure*}[t]
    \includegraphics[width=\textwidth]{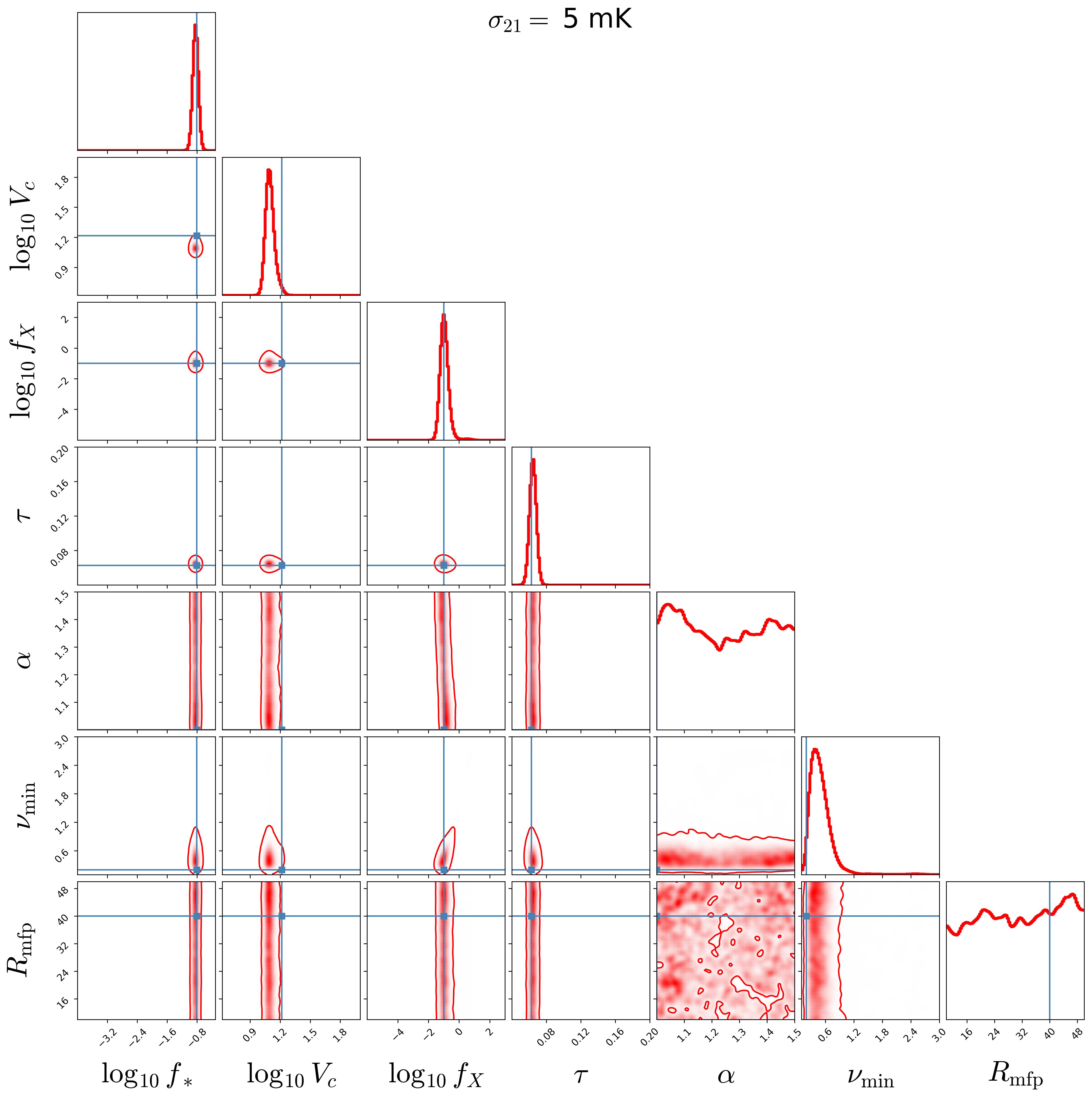}
    \caption{Same as Figure~\ref{fig:appendix_corner_25mK} but when using {\sc 21cmKAN} to fit a different {\sc 21cmGEM} signal with $\theta_0=$($f_*$, $V_c$, $f_X$, $\tau$, $\alpha$, $\nu_{\rm min}$, $R_{\rm mfp}$)=(1.5811388$\times10^{-1}$, 16.5, 0.1, 6.260315$\times10^{-2}$, 1.0, 0.2, 40.0) and the optimistic noise level of $\sigma_{21}=$ 5 mK added (see the left panel of Figure~\ref{fig:posterior_21cmGEM}).\label{fig:appendix_corner_5mK}}
\end{figure*}

\begin{figure*}[t]
    \includegraphics[width=\textwidth]{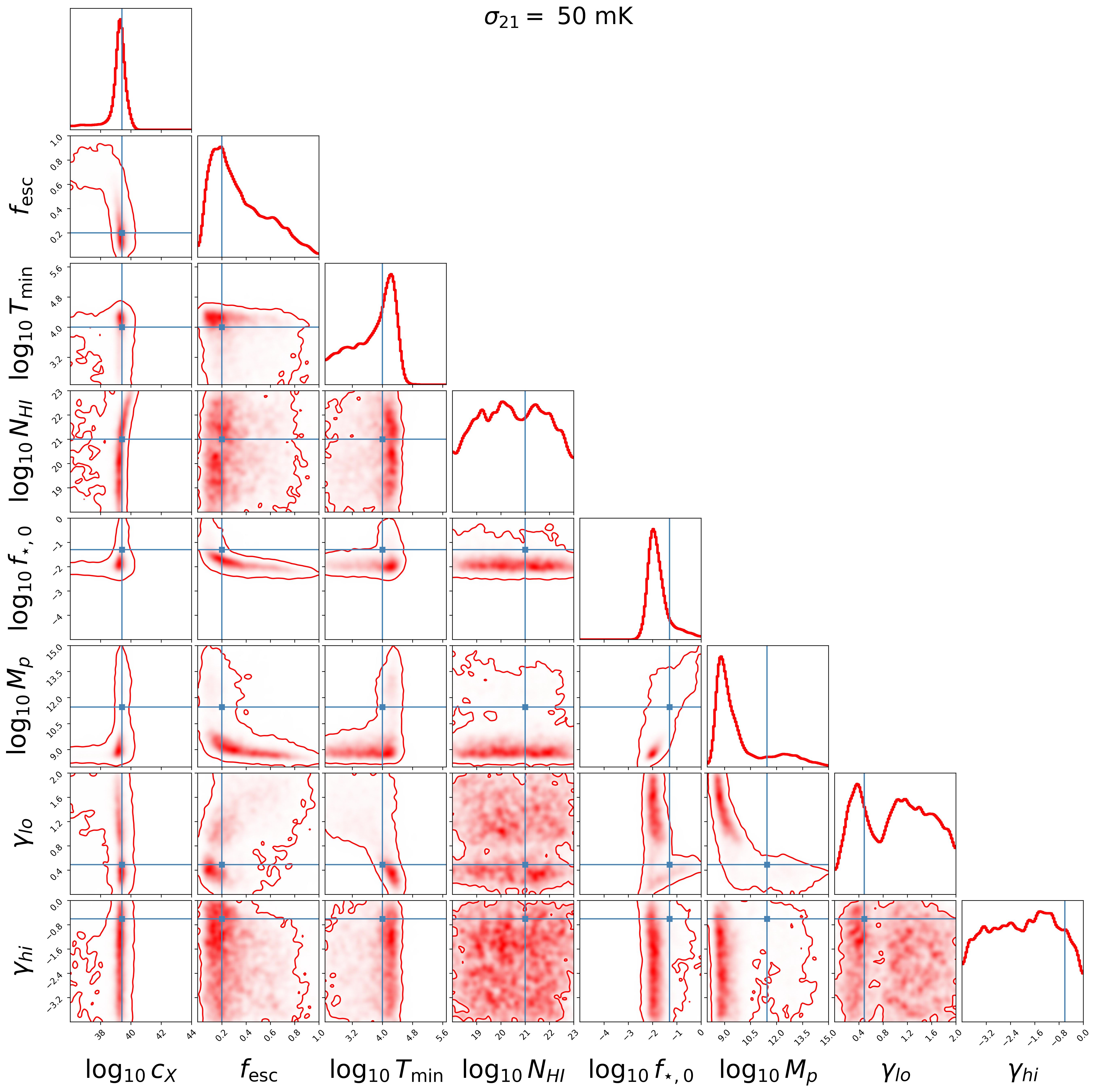}
    \caption{Same as Figure~\ref{fig:appendix_corner_25mK} but when using {\sc 21cmKAN} to fit an {\tt ARES} signal with $\theta_0=$($c_X$, $f_{\rm esc}$, $T_{\rm min}$, $\log N_{\rm H \RomanNumeralCaps{1}}$, $f_{\rm \star,0}$, $M_{\rm p}$, $\gamma_{\rm lo}$, $\gamma_{\rm hi}$)=(2.6$\times10^{39}$, 0.2, $10^4$, 21.0, 0.05, 2.8$\times10^{11}$, 0.49, $-0.61$) and the noise level of $\sigma_{21}=$ 50 mK added. Figure~\ref{fig:posterior_ARES} shows posterior signal realizations for fits to this signal with different $\sigma_{21}$ levels. See Section 2.6 of \citet{DorigoJones23} for a description of how the parameter values of this mock signal were chosen.\label{fig:appendix_corner_50mK}}
\end{figure*}

We present the full 1D and 2D marginalized posterior distributions obtained when using {\sc 21cmKAN} in nested sampling analyses to fit three different mock global 21 cm signals generated by two physical models, {\sc 21cmGEM} and {\tt ARES}, with different added noise levels, $\sigma_{21}$. See Table~\ref{tab:params} for descriptions of the astrophysical parameters being fit, and see Section~\ref{subsec:posterioremulation} and Table~\ref{tab:multinest} for descriptions of the analyses and results and comparison to other works that used different emulators to fit the same signals with the same $\sigma_{21}$. Figures~\ref{fig:appendix_corner_25mK} and~\ref{fig:appendix_corner_5mK} show the posteriors when fitting two different signals from the {\sc 21cmGEM} test set with $\sigma_{21}=$ 25 mK and $\sigma_{21}=$ 5 mK, respectively. Figure~\ref{fig:appendix_corner_50mK} shows the posteriors when fitting the {\tt ARES} signal with $\sigma_{21}=$ 50 mK. The $1\sigma$ posterior signal realizations for the {\sc 21cmGEM} and {\tt ARES} signal fits with $\sigma_{21}=$ 5, 10, and 25 mK are shown in Figures~\ref{fig:posterior_21cmGEM} and~\ref{fig:posterior_ARES}, respectively.

\newpage
\bibliography{dj25}{}
\bibliographystyle{aasjournalv7}
\end{document}